\documentclass[aps,prd,twocolumn]{revtex4-2}

\usepackage[utf8]{inputenc}
\usepackage[table,dvipsnames]{xcolor}
\usepackage{graphicx}
\usepackage{dcolumn}
\usepackage{bm}
\usepackage{amsmath, amssymb, amsthm}
\usepackage{physics}
\usepackage{tikz}
\usetikzlibrary{arrows.meta, positioning}

\usepackage[colorlinks=true,linkcolor=blue,citecolor=blue,urlcolor=blue]{hyperref}

\newcommand{\lrb}[1]{\left( #1 \right)}
\newcommand{\lrrb}[1]{\left[ #1 \right]}

\newcommand{\munu}{\mu\nu}

\begin{document}

	\title{On the Validity of relativistic hydrodynamics beyond local equilibrium} 

	\author{Reghukrishnan Gangadharan}
	\email{reghukrishnang@niser.ac.in}%
	
	\affiliation{School of Physical Sciences, National Institute of Science Education and Research, An OCC of Homi Bhabha National Institute, Jatni-752050, India}

	\date{\today}% It is always \today, today,
	%  but any date may be explicitly specified
	
\begin{abstract}
We examine the applicability of relativistic hydrodynamics far from equilibrium by constructing formal solutions of the Boltzmann moment equations in the relaxation time approximation. These solutions naturally decompose into a divergent gradient series and exponentially decaying non-perturbative modes that encode initial conditions. The non-perturbative contributions are essential for understanding causality, the divergence of the gradient series, and the unexpected effectiveness of relativistic hydrodynamics far from equilibrium.

In the $0~+~1$D Bjorken scenario, we demonstrate that the exact evolution of non-equilibrium terms shares the same structural form as the gradient expansion, differing only through modified transport coefficients that reflect both initial data and free-streaming dynamics. Extending to $3~+~1$D, we find that hydrodynamics remains effective not because the system is close to equilibrium, but because it interpolates smoothly between free streaming and collective behavior. This perspective offers a possible explanation for the remarkable success of hydrodynamics in modeling quark–gluon plasma evolution.
\end{abstract}
	
	%\keywords{Suggested keywords}%Use show keys class option if keyword
	%display desired
	\maketitle

\section{Introduction}

Relativistic hydrodynamics is an effective theory that describes the macroscopic evolution of energy–momentum and conserved charges in a fluid at relativistic energies. It finds key applications ranging from astrophysical plasmas\cite{Bemfica:2019cop,Alford:2017rxf, Shibata:2017jyf}, relativity \cite{Bhattacharyya:2007vjd} and cosmology to high-energy nuclear collisions \cite{Florkowski:2017olj}.

The applicability of relativistic hydrodynamics to the quark–gluon plasma (QGP) is somewhat puzzling. In principle, hydrodynamics is expected to be valid only near local thermal equilibrium, in systems that are sufficiently large and slowly varying compared to microscopic scales. In contrast, the QGP produced in heavy-ion collisions is created in a highly anisotropic, far-from-equilibrium state. Despite these conditions, viscous hydrodynamics provides a quantitatively accurate description\cite{Clare:1986qj} of collective flow observables like particle spectra and flow coefficients \cite{Teaney:2001av,Kolb:2003dz,Teaney:2009qa,Heinz:2024jwu}. Even more surprisingly, flow-like correlations observed in small systems such as $p–Pb$ and high-multiplicity $p–p$ \cite{CMS:2015fgy,ATLAS:2015hzw} collisions can also be reproduced within a hydrodynamic framework. These successes indicate that hydrodynamic behavior can emerge much earlier and in much smaller systems than expected. These observations raise the hydrodynamization question: why does hydrodynamics work so well beyond its domain of validity?

 One possible solution proposed to address this puzzle has been the emergence of hydrodynamic attractors. In highly symmetric settings, solutions of the fluid dynamical equations have been found to converge toward a universal attractor well before the system reaches local equilibrium \cite{Romatschke:2017vte, Heller:2011ju,Heller:2015dha, Buchel:2016cbj,Heller:2013fn}. These attractors have provided useful insights in cases such as Bjorken and Gubser flow, but their extension to non-conformal systems and fully three-dimensional QGP evolution failed to show early time universal behavior \cite{Dash:2020zqx,Chen:2021wwh,Kamata:2022jrc, Jaiswal:2021uvv, Gangadharan:2023yzb}. Consequently, attractors do not provide a full resolution of the applicability of hydrodynamics in heavy-ion collisions.

 An alternate perspective highlights the role of rescaled transport coefficients and fixed points in the effective description \cite{Blaizot:2017ucy,Blaizot:2019scw,Blaizot:2020gql,Blaizot:2021cdv,Jaiswal:2022udf}. It has been demonstrated that second-order hydrodynamic equations can be mapped onto the moment equations of the Boltzmann equation. When these moment equations are truncated at a suitable order, they can reproduce the exact evolution provided the transport coefficients are properly rescaled to align with the fixed points of the exact dynamics.

%%%------------------------------------------------------------------------------------------
 In this work, we present a systematic analytic framework for understanding the emergence of relativistic hydrodynamics from kinetic theory beyond local equilibrium. We demonstrate that the exact solution of the Boltzmann equation, as well as its common model approximations, naturally decomposes into a gradient expansion accompanied by exponentially decaying non-perturbative contributions. In relativistic hydrodynamics, these non-perturbative contributions manifest as the so-called “non-hydrodynamic” modes.  We explore the origin and implications of these non-perturbative terms using singular perturbation theory within the context of a simple toy model.  

We then carry this analysis over to the moment equations by deriving the exact solution of the distribution function moments for a $0~+~1$D system undergoing Bjorken flow.  From this solution, we derive the gradient expansion along with the associated non-perturbative corrections. We obtain evolution equations for the non-equilibrium components of the moments, showing that the dynamical equations obtained using gradient expansion resemble that of the full non-equilibrium dynamics.

Examining the equations for bulk and shear anisotropies, we show that hydrodynamics can be extended to the non-equilibrium free-streaming regime if the transport coefficients are appropriately rescaled. This suggests that conventional relativistic hydrodynamics, with suitably modified transport coefficients, can interpolate smoothly between free-streaming and equilibrium regimes, effectively capturing non-equilibrium effects. We conclude by discussing how these insights extend to systems with more general flow profiles and spacetime dimensions.

The paper is organized as follows. Section 2 reviews the foundations of relativistic hydrodynamics, with emphasis on its derivation from kinetic theory via the Chapman–Enskog expansion. In Section 3, we discuss the breakdown of this expansion and explain the origin of its divergence. Section 5 presents the exact solution of the $0~+~1$D RTA moment equations, from which we derive the corresponding hydrodynamic evolution equations and show how non-equilibrium effects can be incorporated through rescaled transport coefficients. Building on these results, Section 6 extends the analysis to $3~+~1$D relaxation-type equations.

\section{Relativistic Hydrodynamics}
In local thermal equilibrium, the macroscopic state of a system can be described by a finite set of fields: the energy-momentum tensor $T^{\mu\nu}$, the number current $n^{\mu}$, and the fluid four-velocity $u^{\mu}$. At equilibrium these fields can be decomposed as
\begin{align}\label{Eq:EqTnuNmu}
    T^{\mu\nu}_{\text{eq}} = \varepsilon u^{\mu}u^{\nu} - P \Delta^{\mu\nu}\,, &&
    n^{\mu}_{\text{eq}} = nu^{\mu}\,,
\end{align}
where $\varepsilon$ is the fluid rest frame energy density, $P$ the rest frame fluid pressure, $n$ the number density and the four-velocity is normalized to $u^{\mu}u_{\mu} = 1$. The tensor $\Delta^{\mu\nu} = g^{\mu\nu} - u^{\mu}u^{\nu}$ acts as a projection operator that maps any vector onto the subspace orthogonal to $u^{\mu}$.  The conservation equations 
\begin{align}\label{Eq:Cons}
    \partial_{\mu}T^{\mu\nu}  &= 0\,,\\
    \partial_{\mu}n^{\mu}~~ &=0\,,
\end{align}
together with an equation of state of the form $\varepsilon = \varepsilon(n,P)$, provide a closed set of equations for describing the evolution of the system in local equilibrium. Alternatively, one may describe the equilibrium state using temperature ($T$) and chemical potential($\mu$) instead of $\varepsilon$, $n$, and $P$.

Away from equilibrium, the energy-momentum tensor and the particle current are typically expressed as a sum of equilibrium and dissipative contributions:
\begin{align}
   T^{\mu\nu}  = T^{\mu\nu}_{\text{eq}}  + \Pi^{\mu\nu}\,, &&
    n^{\mu} =n^{\mu}_{\text{eq}} + q^{\mu}\,.
\end{align}
However, this decomposition is not uniquely defined in the out-of-equilibrium case, since the fluid four-velocity lacks a unique definition. Common conventions for fixing the velocity include the Landau frame and the Eckart frame. In the Landau frame, there is no energy dissipation, whereas in the Eckart frame, particle diffusion is absent.  Throughout this paper, when a frame choice is required, we will adopt the Landau frame defined by,
\begin{align} \label{Eq:LandauFrame}
    T^{\mu\nu}u_{\nu} = \varepsilon u^{\mu} \,, &&
     n^{\mu}u_{\mu}  = n \,.
\end{align}
The dissipative part of the energy-momentum tensor, $\Pi^{\mu\nu}$, is further split into a scalar (trace) and a traceless symmetric tensor,
\begin{align}
    \Pi^{\mu\nu} = -\Pi \Delta^{\mu\nu} + \pi^{\mu\nu}\,,
\end{align}
where $\Pi$ denotes the bulk viscous pressure, and $\pi^{\mu\nu}$ is the shear pressure tensor. In the non-relativistic case these components are expanded in the gradients of $u^{\mu}$. Temporal gradients of the energy density (or temperature $T$) and particle density (or chemical potential $\mu$) can be eliminated using the conservation equations \cite{Jaiswal:2014isa}. A relativistic generalization of this gradient expansion up to first order gives the relations,
\begin{align}\label{Eq:HydFirst}
    \pi^{\mu\nu}    = 2\eta\sigma^{\mu\nu} \,, &&
    \Pi             = -\zeta\theta\,,
\end{align}
where $\eta$ and $\zeta$ are the shear and bulk viscosity, respectively.  However, it is well known that the first-order gradient expansion leads to acausal behaviour \cite{Hiscock:1987zz,Hiscock:1986zz,Denicol:2008ha}.

To ensure causality, the anisotropic stresses $\pi^{\mu\nu}$ and $\Pi$ are treated as independent dynamical fields with their own evolution equations. In second-order Israel-Stuart \cite{Israel:1976tn,Israel:1979wp} like theories, these typically take the form:
\begin{align}\label{Eq:HydSec}
    \dot{\Pi} &= -\frac{\Pi}{\tau_{\Pi}} -\frac{\zeta}{\tau_{\Pi}}\theta  &+~\text{additional terms}\,,\\
    \dot{\pi}^{\mu\nu} &= -\frac{\pi^{\mu\nu}}{\tau_{\pi}} + 2\frac{\eta}{\tau_{\pi}}\sigma^{\mu\nu} &+~ \text{additional terms}\,,
\end{align}
where $\theta \equiv \partial_{\mu}u^{\mu}$ is the expansion scalar and $\sigma^{\mu\nu} \equiv \frac{1}{2}(\nabla^{\mu}u^{\nu}+ \nabla^{\nu}u^{\mu}) - \frac{1}{3}\theta \Delta^{\mu\nu}$ is the shear tensor. The commoving derivatives are given by $\dot{A} = u^{\mu}\partial_{\mu}A$ (temporal) and $\nabla^{\mu} = \Delta^{\mu\nu}\partial_{\nu}$( spatial ). The additional terms encapsulate higher-order corrections and possible couplings to other dynamical variables. Here, the relaxation times $\tau_{\pi}$ and $\tau_{\Pi}$ control how quickly the system approaches equilibrium, ensuring this occurs over a finite timescale. 

A systematic way to derive these dynamical equations is to begin with the evolution equations of an underlying microscopic theory and use the connection between macroscopic moments and microscopic degrees of freedom to obtain their dynamics. The relativistic Boltzmann equation serves as a natural choice for such a microscopic description.

\subsection{Kinetic theory to hydrodynamics}

The Boltzmann equation \cite{Cercignani:2002book},
 \begin{align}
     \lrb{p^{\mu}\pdv{}{x^{\mu}} + mK^{\mu}\pdv{}{p^{\mu}} - \Gamma^{\sigma}_{\munu}p^{\mu}p^{\nu}\pdv{}{p^{\sigma}}}f(x^{\mu},p^{\mu}) = C[f,f]\,, \label{Eq:BolF}
 \end{align}
 governs the evolution of the single particle phase space distribution function $f(x^{\mu},p^{\mu})$ for systems in the dilute limit, where the mean free path $\lambda$ is much larger than the characteristic interaction length scale. For simplicity, we will often omit the explicit dependence of $f$ on its variables, with the understanding that it is generally a function of both $x^\mu$ and $p^\mu$. Constraints on the spacetime and momentum dependence of $f$ arising from symmetries will be discussed in the relevant sections.
 
 In this expression, $K^{\mu}$ represents an external four-force, and $\Gamma^{\sigma}_{\mu\nu}$ are the Christoffel symbols accounting for spacetime curvature. The four-momentum $p^{\mu}$ obeys the on-shell condition $g_{\mu\nu}p^{\mu}p^{\nu} = m^2$, while the force term satisfies $g_{\mu\nu}p^{\mu}K^{\nu}/p^0 = 0$. On the right-hand side, we have the collision kernel $C[f,f]$ defined as,
 \begin{equation}
C[f,f] = \int dP_1dP'dP_1'
~W(p, p_1 \mid p', p_1') \left[ f' f_1' - f f_1 \right]
\end{equation}
where $W(p, p_1 \mid p', p_1')$ is the Lorentz invariant transition rate for the process $p+p_1 \to p' + p_1'$ and $dP = d^3p/(2\pi)^3 2p^0$. The term in brackets, $[f’ f_1’ - f f_1] $ represents gain and loss due to collisions. The vanishing of the collision kernel defines local thermal equilibrium distribution $C[f_{eq},f_{eq}] = 0$, 
 \begin{equation}
    f_{\text{eq}} = \frac{1}{e^{\lrb{u\cdot p - \mu}/T} + r},
\end{equation}
where $r$ takes values ${-1,0,1} $ corresponding to Bose-Einstein, Maxwell-Juttner, and Fermi-Dirac statistics, respectively.

The hydrodynamic fields are obtained as momentum integrals of the distribution function,
 \begin{align}
    n^{\mu}         = \int \frac{d^3p}{2\pi^2 p^0} p^{\mu}f \,,&&
     T^{\mu\nu} = \int \frac{d^3p}{2\pi^2 p^0} p^{\mu}p^{\nu}f\,.
 \end{align}
To separate the equilibrium and non-equilibrium contributions to these fields, the distribution function is decomposed into an equilibrium part and a deviation from equilibrium,
 \begin{align}
     f = f_{\text{eq}} (1 + \phi)\,.
 \end{align}
 where 
 \begin{equation}\label{Eq:df}
     \delta f \equiv  (f - f_{\text{eq}}) = f_{\text{eq}}\phi\,.
 \end{equation}

 The non-equilibrium contributions to the hydrodynamic fields can then be expressed in terms of $\delta f$,
\begin{align}
    q^{\mu}         = \int \frac{d^3p}{2\pi^2 p^0} p^{\mu}\delta f \,,&&
     \Pi^{\mu\nu} = \int \frac{d^3p}{2\pi^2 p^0} p^{\mu}p^{\nu}\delta f\,.
     \label{Eq:AnisoBol}
 \end{align}
The temperature $T$, chemical potential $\mu$, and fluid four-velocity $u^{\mu}$ that 
appear in the equilibrium distribution function are determined using the Landau frame conditions, given in Eq.\eqref{Eq:LandauFrame}. The evolution equations for the
anisotropies can readily be obtained by substituting \eqref{Eq:AnisoBol} in \eqref{Eq:BolF}. However, these equations are of limited practical use, as they require knowledge of the full distribution function $f$ or its infinite hierarchy of moments. This motivates the need for an approximation scheme along with a suitable closure prescription.

\subsection{The Chapman-Enskog expansion}

The Chapman-Enskog expansion assumes that the distribution function depends functionally on the fluid-dynamical variables—such as energy density $\varepsilon$, number density $n$,  anisotropy $\Pi^{\mu\nu}$ and their gradients- $ f(\rho,\nabla \rho)$ or more generally, a hierarchy of moments $\rho_n$)—and that this dependence can be systematically expanded in powers of a small parameter $Kn \sim \lambda/L$,
\begin{align}
    f &= f^{(0)} + Kn~f^{(1)} + Kn^{2}~f^{(2)} \dots \,,\nonumber\\
    \rho_{n} & = \rho_{n}^{(0)} + Kn~\rho_{n}^{(1)} + Kn^{2} ~\rho_{n}^{(2)} \dots\,,
\end{align}
where $\lambda$ denotes the microscopic mean free path and $L$ represents a characteristic macroscopic length scale. This approach is valid when there is a clear separation between microscopic and macroscopic scales - the former governs the relaxation of the distribution function toward the local Maxwell-Juttner equilibrium, while the latter controls the emergence of viscous hydrodynamic behavior \cite{Grad:1958}. The Chapman–Enskog expansion for the Boltzmann collision kernel is discussed in detail in \cite{chapman:1990}. 

In this section, we demonstrate the expansion scheme using the Boltzmann equation with the Anderson–Witting \cite{Anderson:1974nyl} relaxation-time approximation (AW-RTA),
\begin{equation}\label{Eq:BolRTA}
p^{\mu} \partial_{\mu} f = -\frac{u \cdot p}{\tau_R}(f - f_{\text{eq}}),,
\end{equation}
where $\tau_R$ is a model parameter representing the microscopic relaxation time. Although equation \eqref{Eq:BolRTA} takes the form of a linear inhomogeneous partial differential equation, it is effectively nonlinear. This is because the equilibrium distribution function $f_{\text{eq}}$ depends on the temperature $T$, chemical potential $\mu$, and fluid velocity $u^\mu$, which are themselves determined through the Landau matching conditions \eqref{Eq:LandauFrame}. As a result, $f_{\text{eq}}$ has a non-trivial dependence on $f$.

Here we can identify $\frac{\tau_R}{u\cdot p} p^{\mu}\partial_{\mu} \sim \frac{\tau_R}{L} \sim Kn$. An iteration on \eqref{Eq:BolRTA} gives a formal expansion in powers of Knudsen number,
\begin{equation}\label{Eq:GradCE}
    f = f_{eq} + \sum_{k=1}^{\infty}\lrrb{\frac{\tau_R}{u\cdot p}p^{\mu}\partial_{\mu}}^{k} f_{eq}\,.
\end{equation}
Note that the above expression gives $f$ completely in terms of $f_{eq}$. Therefore, in principle, all anisotropic moments in Eq.~\eqref{Eq:AnisoBol} can be expressed in terms of the equilibrium fields $T$, $\mu$, $u^{\mu}$, and their derivatives. To treat the anisotropies as independent dynamical fields, however, one additional step is necessary. We obtain evolution equations for the dissipative quantities via  

\begin{align}
    \dot{q}^{\mu}         = \int \frac{d^3p}{2\pi^2 p^0} p^{\mu}\delta \dot{f} \,,&&
     \dot{\Pi}^{\mu\nu} = \int \frac{d^3p}{2\pi^2 p^0} p^{\mu}p^{\nu}\delta \dot{f}\,,
     \label{Eq:AnisoBolDyn}
 \end{align}
 from \eqref{Eq:GradCE}. The expansion is truncated at a chosen order $n$ \cite{Denicol:2010xn,Denicol:2012cn,Jaiswal:2013npa,Jaiswal:2013vta,Panda:2020zhr,Panda:2021pvq}. 
 
 %$\sigma^{\mu\nu} \to \frac{\pi^{\mu\nu}}{2\eta}$ and $\theta \to \tfrac{\Pi}{\zeta}$ made at each order . 
 
 % {\color{Blue} See if its done at every order}{\color{Red} This effectively acts as a re-summation of the gradient expansion\cite{Romatschke:2017ejr}.  } 

\section{Divergence of the Gradient Expansion}

The Chapman–Enskog expansion is an asymptotic, and generally divergent, solution of the Boltzmann equation \cite{Grad:1963,Heller:2013fn,Buchel:2016cbj,Denicol:2016bjh,Heller:2015dha}. The divergence of the Chapman–Enskog expansion can be understood using the framework of singular perturbation theory. To illustrate this, consider the toy differential equation
\begin{equation}\label{Eq:ToyModel}
\delta\frac{df(t)}{dt} = -\lrb{f(t) - g(t)}\,,
\end{equation}
where $\delta$ is a small parameter. Despite its simplicity, the toy model captures several essential features of the Boltzmann equation and its hydrodynamic limit.

In \eqref{Eq:ToyModel} the highest derivative term is multiplied by the perturbative factor $\delta$. A straightforward perturbative expansion of $f(t)$ in powers of $\delta $ yields the solution,
\begin{equation}\label{Eq:ToyModelGExp}
f(t) = g(t) - \delta~ g'(t) + \delta^2~ g''(t) + \dots \,.
\end{equation}
This solution fails to accommodate arbitrary initial conditions for $f(t)$, and diverges when $\delta \gtrsim 1$, unless $g(t)$ is a polynomial. By contrast, the exact solution
\begin{equation}\label{Eq:ToyExSol}
f(t) = e^{-t/\delta}f(0) + \int_{0}^{t}\frac{dt'}{\delta}e^{-(t-t')/\delta}g(t’)
\end{equation}
is free of both issues, it correctly incorporates initial data and remains well-behaved for all values of $\delta$. Performing integration by parts on the second term of \eqref{Eq:ToyExSol} yields,
\begin{align}\label{Eq:ToyGradInt}
     f(t) &= e^{-t/\delta}f(0) + \lrb{g(t)- e^{-t/\delta}g(0) }\nonumber\\
     &- \delta \lrb{~g'(t)- e^{-t/\delta} g'(0)} +\dots \,.
\end{align}
This expression matches the perturbative series in the late-time limit, $t \gg \delta$ irrespective of initial conditions. But the expansion around $\delta = 0$ misses contributions that are non-analytic or of order $1/\delta$, and breaks down when $t \sim \delta$. 
If $g(t)$ is analytic in $t$, Taylor expanding $g(t)$ and integrating gives an alternate (equivalent) solution for \eqref{Eq:ToyExSol},
\begin{equation}\label{Eq:ToyGradTay}
    f(t) = e^{-t/\delta}f(0) + \sum_{n=0}^{\infty}\lrb{1 - e^{-w}\sum_{k=0}^{n}\frac{w^k}{k!}}\frac{d^n}{dt^n}g(t)\,,
\end{equation}
where $w = (t-t_0)/\delta$ and $\gamma(n+1,w)$ is the lower incomplete gamma function defined as,
\begin{equation}
    \frac{\gamma(n+1,w)}{n!} = \lrb{1 - e^{-w}\sum_{k=0}^{n}\frac{w^k}{k!}}\,.
\end{equation}

This expression is advantageous because it represents the solution primarily through local derivatives, each weighted by time-dependent coefficients. As a result, when the pre-factor $e^{-t/\delta}f(0) \ll 1$, the solution is well approximated by the series terms alone.
\subsubsection*{The Fixed point perspective}
The two limits of the exact solution to \eqref{Eq:ToyModel} can be understood in terms of its fixed points (where the derivative vanishes). The equation has an attractive fixed point at $\delta \to 0$ ($f = g$) and a repulsive fixed point at $\delta \to \infty$ ($\dot{f} = 0$). A perturbative expansion around the attractive fixed point $\delta = 0$ yields the derivative series \eqref{Eq:ToyModelGExp}, while an expansion around $1/\delta = 0$ captures the exponentially decaying non-perturbative corrections. The exact solution smoothly interpolates between the behaviors near these two fixed points. As we will see, the applicability of hydrodynamics stems from its ability to interpolate between the two fixed points of the Boltzmann equation.

\subsection{The Boltzmann Kernel}
 For the full Boltzmann collision kernel, if the characteristic interaction range is much smaller than the mean free path, one can separate the incoming and outgoing contributions to the collision operator and the Boltzmann equation can be cast in the form,
\begin{equation}\label{Eq:BolSep}
v^{\mu}\partial_{\mu}f +\frac{v}{\delta}f = \frac{1}{\delta}L[f],,
\end{equation}
where $\text{v}^{\mu} = p^{\mu}/p^{0}$ is the four velocity of the particle and $v(p) \sim 1/\tau_R$ is the average collision frequency, 
\begin{equation}
\nu(p) = \int dP_1dP'dP_1 \,
W(p, p_1 \mid p', p_1') \, f(p_1)\,.
\end{equation}

and 
\begin{equation}
L[f] = \int dP'  dP_1' dP_1 \; W(p', p_1' \mid p, p_1) \; f' f_1'
\end{equation}
 A direct integration of \eqref{Eq:BolSep} using the method of characteristics yields a solution \cite{Grad:1958} (subject to appropriate boundary conditions),
\begin{align}\label{Eq:BolExSol}
f(p^{\mu},x^{\mu}) = &e^{-\frac{1}{\delta}\int_{t_0}^{t} v(t,t')}f(t,t_0) \nonumber\\
&+\int_{t_0}^{t}  \frac{v(t,t')}{\delta} e^{-\frac{1}{\delta}\int_{t'}^{t} v(t,t'') dt''} g(t,t') dt'
\end{align}
where $g = L/v$. Here we have introduced the shorthand $h(t,t’) = h(p^\mu, x^{\mu} - (p^{\mu}/p^{0})(t - t’))$ to represent the free-streamed value of any function $h(p^{\mu},x^{\mu})$. This solution is structurally similar to that obtained for the relaxation time approximation \cite{Gangadharan:2024ovs}. Following the same procedure in \cite{Gangadharan:2024ovs}, we can define $\chi' = \frac{1}{\delta}\int_{t'}^{t}v dt'$, we can obtain a formal series expansion for the full solution to the Boltzmann equation,
\begin{align}\label{Eq:GrdSerBol}
f &= e^{-\chi_0}f(t,t_0) \nonumber\\
&+ \sum_{n=0}^{\infty} \delta^n \frac{\gamma(n+1,\chi_0)}{n!} \lrrb{-\frac{p^{\mu}\partial_{\mu}}{v}~  }^{n}g\,.
\end{align}

A convergent series expansion was constructed \cite{Gangadharan:2024ovs} for the relaxation time approximation (RTA) kernel \eqref{Eq:BolRTA},
\begin{align}\label{Eq:GrdSerRTA}
f &= e^{-\xi_0}f^0(t,t_0) \nonumber\\
&+ \sum_{n=0}^{\infty} \frac{\gamma(n+1,\xi_0)}{n!} \lrrb{-\frac{\tau_R}{p \cdot u} p^{\mu}\partial_{\mu}}^{n} f_{\text{eq}}\,,
\end{align}
where $\xi_0 = \int_{t_0}^{t} \frac{u \cdot p}{\tau_R} dt’$ is the damping factor. Notably, it mirrors the solution \eqref{Eq:GrdSerBol} under the identifications $v(p) \sim u\cdot p/\tau_R $, $\chi_0 \sim \xi_0$ and $g \sim f_{\text{eq}}$. In fact any relaxation type model - $L[f] \to v(p)g[f]$ for the collision kernel will have a gradient  expansion structurally similar to \eqref{Eq:GrdSerBol}.

\subsection{Validity of the Gradient Expansion}

Before discussing the validity of the relativistic gradient expansion, it is important to distinguish between the gradient expansion itself and relativistic hydrodynamics. The gradient expansion refers specifically to the asymptotic series obtained via a perturbative expansion in the Knudsen number ($Kn$). Relativistic hydrodynamics, on the other hand, encompasses not only this gradient series but also non-perturbative corrections—an aspect that will be made more clear in later sections. The expansions shown in \eqref{Eq:GrdSerBol} and \eqref{Eq:GrdSerRTA} naturally include non-perturbative contributions—often labeled as “non-hydrodynamic” modes. This label, however, can be misleading as such modes are essential for ensuring both causality \cite{Gavassino:2023mad} and convergence \cite{Gangadharan:2024ovs} of the hydrodynamic theory. It is more accurate to interpret them as non-perturbative corrections, not non-hydrodynamic.

The exponentially decaying non-perturbative terms in \eqref{Eq:GrdSerRTA} can be rewritten \cite{Gangadharan:2024ovs} in the form,
\begin{equation}
    f_{NP} =- e^{-\xi'(t,t')}\lrrb{-\frac{\tau_R(t,t_0) }{ p \cdot U(t,t_0) } p^{\mu}\partial_{\mu}}^{n}  f_{eq}( t,t_0)\,.
\end{equation}

It is clear that these non-perturbative contributions encode information about the initial conditions and become relevant when the Knudsen number approaches unity ($Kn \sim 1$)—that is, during early times when $\xi_0 < 1$, or equivalently, when $t < \tau_R$ for constant relaxation time. Unlike the toy model, these corrections are non-local due to the dependence of $\tau_R$ on the evolving moments of the distribution function. Physically, this regime corresponds to the pre-collisional or free-streaming phase, where interactions have yet to dominate. In this limit, the traditional gradient expansion breaks down because it cannot fully capture the free streaming structure of the distribution function.

For a smooth, well-posed dynamical system, fixed points can only be approached asymptotically and never reached at finite time. The Chapman–Enskog gradient expansion, however, implicitly assumes that one can construct a solution valid at every finite time by expanding around the fixed point (the local equilibrium). 

The Chapman-Enskog gradient expansion remains valid as long as free-streaming effects are suppressed—i.e., when $\chi_0$ or $\xi_0 \gg 1$—even when the gradients are large.
We therefore argue that the applicability of the gradient expansion should not be judged by small gradients alone, but rather by how far ($\chi_0~\text{or}~\xi_0 \gg 1$) it is from the free-streaming regime. In the simple one-dimensional toy model these two conditions are equivalent, since $t/\delta \gg 1 \implies \delta,\frac{d}{dt} \ll 1$, but this equivalence does not necessarily hold in more general cases.

This analysis makes explicit that the standard gradient expansion captures only the asymptotic behavior near the attractive fixed point, while missing the exponentially suppressed contributions that encode initial conditions. In the following sections, we show that these non-perturbative contributions are not merely correction, but play a central role in extending the validity of hydrodynamics far from equilibrium.

\section{Hydrodynamics in ${\mathbf0+1}$ D}
In this section, we examine the applicability of hydrodynamics by studying a simple $0~+~1$D system, using it as a foundation for more general insights. We begin by deriving the moment equations of the Boltzmann equation in the relaxation time approximation (RTA). Exploiting the linearity of the RTA kernel, we construct formal solutions for these moments. These solutions are then used to obtain the exact evolution equations for hydrodynamic variables. Finally, we compare these exact results with those predicted by second-order hydrodynamics, enabling us to assess how well hydrodynamics approximates the true evolution and to delineate its regime of validity.

We consider a system with rotational and translational symmetry in the x\text{–}y plane and boost invariance along the z-axis. It is then convenient to work in the following Milne coordinate system, 
\begin{equation}
    \begin{aligned}
    \tau &= \sqrt{t^2-z^2}\,,\\
    \eta &= \tanh^{-1} (z/t)\,,\\
    r &=\sqrt{x^2=y^2}\,,\\
    \theta&= \tan^{-1}(y/x)\,.  
\end{aligned}
\end{equation}
These symmetries restrict the spacetime dependence to the proper time $\tau$ alone, thus $0~+~1$D. The fluid four-velocity in Bjorken flow \cite{Bjorken:1982qr} is given in Cartesian coordinates by $u^{\mu} = (\cosh{\eta}, 0, 0, \sinh{\eta})$. When transformed to Milne coordinates, this simplifies to $u^{\mu} = (1, 0, 0, 0)$. In this coordinate system, the metric takes the form $g_{\mu\nu} = \text{diag}(1, -1, -r^2, -\tau^2)$ and the only non-vanishing component of the velocity gradient is $\nabla_{\mu} u^{\mu} = \frac{1}{\tau}$.

The symmetries put strong constraints on the form of the energy-momentum tensor. In Milne coordinates $(\tau, x, y, \eta_s)$ must take the diagonal form,
\begin{equation}
    T^{\mu\nu} =
\begin{pmatrix}
\varepsilon & 0 & 0 & 0 \\
0 & P + \Pi+\frac{\pi}{2}& 0 & 0 \\
0 & 0 & P+\Pi +\frac{\pi}{2}& 0 \\
0 & 0 & 0 & P+\Pi -\pi
\end{pmatrix} \,,
\end{equation}
where $\varepsilon$ is the energy density, $P$ the equilibrium isotropic pressure, $\Pi$ is the bulk pressure and the shear tensor is reduced to the form $\pi^{\mu\nu} =  \text{diag}(0,\frac{\pi}{2},\frac{\pi}{2},-\pi)$. The transverse pressure $P_T = P+\Pi +\frac{\pi}{2}$ and the longitudinal pressure is $P_L = P+\Pi -\pi$.

The RTA Boltzmann  equation in for this system takes the simplified form,
\begin{equation}\label{Eq:ZoRTABoleq}
\pdv{f}{\tau}- \frac{p^{z}}{\tau}\frac{\partial f}{\partial p^{z}}= -\frac{1}{\tau_R}(f - f^{\rm eq})\,.
\end{equation}
The distribution function $f(p^0,p^z,\tau)$ depend on the particle energy $p^0$, the $z$ component of momentum $p^z$ and the proper time $\tau$. The symmetries in the transverse plane imply transverse momentum space isotropy.  The transverse momentum is given by the energy-momentum relation $(p^0)^2 - (p^z)^2 -p_T^2 = m^2$, where $m$ is the mass of the particle and $p_T = \sqrt{(p^x)^2 + (p^y)^2}$. 

\subsection{Moment Equations}
We define \cite{Denicol:2016bjh} the moments $\rho_{n,l}$ of the distribution function as
\begin{equation}\label{Eq:ZoMomDef}
    \rho_{n,l}(\tau) = \int \frac{d^{3}p }{(2\pi)^3 p^{0}}(p^{0})^{n+1}\lrb{\frac{p^{z}}{p^{0}}}^{2l} f(p^0,p^z,\tau)\,.
\end{equation}
Here, the index $n$ encodes the energy scaling, while $l$ characterizes the anisotropy in the momentum distribution. In particular, the number density ($n$) corresponds to the moment $\rho_{0,0}$, and the energy density ($\varepsilon$) is given by $\rho_{1,0}$. The equilibrium moments $\rho_{n,l}^{\rm eq}$ when $f$ has the Maxwell-Juttner form has the expression,
\begin{equation}\label{Eq:ZoEqlMom}
   \rho_{n,l}^{\rm eq}(T,z) =   \frac{T^{n+3}}{2\pi^2}\frac{G_{n+3,l}(z) }{2l + 1}  e^{\mu/T}  \,,
\end{equation}
where $G_{n,l}(z)$ is a function (Appendix \eqref{Ap:GDef}) that depends on the reduced mass $z = m/T$. In the massless limit  $\lim_{z \to 0} G_{n,l}(z) = \Gamma(n)$. 

The dynamics of the moments in $0~+~1$D  can be obtained \cite{Denicol:2016bjh} by integrating the RTA equation \eqref{Eq:ZoRTABoleq}  ,
\begin{equation}\label{Eq:ZoRTAMomeq}
    \partial_{\tau}\rho_{n,l} + \frac{2l+1}{\tau}\rho_{n,l} - \frac{2l-n}{\tau}\rho_{n,l+1}= -\frac{1}{\tau_{R}}\lrb{\rho_{n,l} -\rho_{n,l}^{\text{eq}}}\,.
\end{equation}

In the moment equations Eq.\eqref{Eq:ZoRTAMomeq}, the $l$th moment is coupled with the $l+1$th moment, forming an infinite set of coupled differential equations. These equations are typically solved numerically by using a truncation scheme where all moments for $l>l_m$ is set to zero \cite{deBrito:2023vzv,Denicol:2016bjh,Jaiswal:2022udf,Gangadharan:2025qbp} for some arbitrary choice of $l_m$. 

\subsection{Solution of the moment equations}
 A set of coupled linear equations can always be written in a matrix form. Though \eqref{Eq:ZoRTAMomeq} is a nonlinear differential equation since $\tau_R$ generally depends on temperature, it has a linear algebraic structure. Exploiting this we can write \eqref{Eq:ZoRTAMomeq} in the form,
 \begin{equation}\label{Eq:ZoRTAMOmOp}
     \lrrb{\partial_{\tau} + \frac{2l+1}{\tau}\mathbf{\hat{I}} - \frac{2l-n}{\tau}\mathbf{\hat{S}} }\rho_{n,l}= -\frac{1}{\tau_{R}}\mathbf{\hat{I}}\lrb{\rho_{n,l} -\rho_{n,l}^{\text{eq}}}\,,
 \end{equation}
 where we defined the identity operator $\mathbf{\hat{I}}$ and the $l$ shift operator,
 \begin{align}
     \mathbf{\hat{S}}[h(l)] = h(l+1)\,.
 \end{align}
 For reasons that will become clear later we define the operator,
 \begin{align}
     \mathbf{\hat{F}}(\tau) &= \frac{2l+1}{\tau}\mathbf{\hat{I}} - \frac{2l-n}{\tau}\mathbf{\hat{S}}\,,\\
     \mathbf{\hat{D}}(\tau) &= \partial_{\tau} +  \mathbf{\hat{F}}(\tau)\,
 \end{align}
The operator $\mathbf{\hat{F}}$ can be interpreted as a Liouville-like operator acting in moment space.  It is now instructive to look at the solutions to the collision free equation,
 \begin{align}
     \lrrb{ \partial_{\tau} + \mathbf{\hat{F}}(\tau) }g_{n,l}(\tau)  = 0\,.
 \end{align}
Using the integrating factor method we can write down the solution (Appendix \eqref{Ap:ZpOD}),
 \begin{equation}
     g_{n,l}(\tau)  = e^{-\mathbf{\hat{K}(\tau,\tau_0)}}g_{n,l}(\tau_0) \,
 \end{equation}
 where $\mathbf{\hat{K}}(\tau,\tau_0)\equiv \int_{\tau_0}^{\tau} \mathbf{\hat{F}}(\tau') d\tau'$. If the initial values $g_{n,l}(\tau_0)$ are the moments of a distribution function $f$, then the evolved quantities $g_{n,l}(\tau)$ represent the moments of the free-streamed distribution function $f(\tau,\tau_0)$.
 
 We can interpret this more elegantly if we consider the vector space spanned by the moments $\rho_{n,l}(\tau)$. Then the operator $\mathbf{\hat{F}}(\tau')$ is the generator of free streaming group with the representation (propagator) $\hat{\mathcal{F}}_{\tau,\tau'} = e^{-\mathbf{\hat{K}}(\tau,\tau')}$.

An exact expression for the  free streaming solution can be obtained (see Appendix \eqref{Ap:ZpOD})
 \begin{equation}\label{Eq:FSExSol}
     g_{n,l}(\tau) =  \sum_{k=0}^{\infty}\, K_{n,l,k}(\tau,\tau_{0})g_{n,l+k}(\tau_0) \,,
 \end{equation}
 where
\begin{equation}\label{Eq:GenInt_Ker}
    K_{n,l,k}(\tau,\tau^{\prime}) =  \frac{\lrb{l -\frac{n}{2}}^{(k)}}{k!} \lrb{\frac{\tau^{\prime}}{\tau}}^{2l+1}\lrrb{1 - \lrb{\frac{\tau'}{\tau}}^{2}}^{k} \,.
\end{equation}
with $(\cdot)^{(k)}$ being the rising Pochhammer symbol.

The leading order late time behavior can be easily read off from the expression $g_{n,l}(\tau>>\tau_0) \sim \lrb{\frac{\tau_0}{\tau}}^{2l+1}$, which is independent of $n$. 

% \begin{figure}
%     \centering
%     \includegraphics[width=1.0\linewidth]{FS_AvsN.pdf}
%     \caption{The $\log-\log$ plot of the free streaming of the moment $\rho_{1,1}$ as a function of time for the analytical solution (dotted) obtained in \eqref{Eq:FSExSol} vs the numerical solution (solid) for the moment equation. }
%     \label{fig:NumVsAnl}
% \end{figure}

We can now write down a formal solution to the RTA moment equation, 
\begin{align}\label{Eq:MomFormSol}
    \rho_{n,l}(\tau) &= e^{-\xi_0}e^{-\mathbf{\hat{K}}(\tau,\tau_0)}\rho_{n,l}^{0}(\tau) \nonumber\\
    & + \int_{\tau_0}^{\tau}d\tau^{\prime} \frac{e^{-\xi'}}{\tau_{R}}e^{-\mathbf{\hat{K}}(\tau,\tau')}\rho_{n,l}^{\text{eq}}(\tau^{\prime})\,,
\end{align}
for the moment equation \eqref{Eq:ZoRTAMOmOp}, where $\xi' = \int_{\tau'}^{\tau}d\tau''/\tau_R$. The similarity of \eqref{Eq:MomFormSol} to \eqref{Eq:ToyExSol} and \eqref{Eq:BolExSol} is immediately obvious.  The integral term in the above expression $\rho_G$, called the hydrodynamic generator in \cite{McNelis:2020jrn} can be expanded as ( Appendix \eqref{Ap:ZpOD})
\begin{align}\label{Eq:GenIntSer_G}
    \rho^{G}_{n,l}(\tau) =  \,\int_{\tau_0}^{\tau}d\tau^{\prime}\frac{e^{-\xi'}}{\tau_{R}} \sum_{k=0}^{\infty}K_{n,l,k}(\tau,\tau^{\prime}) \rho_{n,l+k}^{\text{eq}}(\tau^{\prime})\,.
\end{align}
In the conformal case ($m=0$), this expression simplifies ( Appendix \eqref{Ap:ZpOD}) to the known closed-form result in terms of hypergeometric functions \cite{Strickland:2019hff}. Since the hypergeometric function is absolutely convergent in the relevant domain, the order of summation and integration can be interchanged. This further confirms that the exact gradient expansion derived from this expression is convergent.

\subsection{Gradient expansion of the moments}
A gradient expansion for the moments can be obtained( Appendix \eqref{Ap:ZpODGrd}) from the solution \eqref{Eq:MomFormSol},
\begin{align}\label{Eq:MomFormSolTyl}
   \rho_{n,l}(\tau)  &= e^{-\xi_0}e^{-\mathbf{\hat{K}}(\tau,\tau_0)}\rho_{n,l}(\tau_0) \nonumber\\
   &+  \sum_{k=0}^{\infty} \frac{\gamma(k+1,\xi_0)}{k!} \lrrb{-\tau_R \mathbf{\hat{D}}  }^{k} \rho^{\text{eq}}_{n,l}(\tau)\,.
\end{align}
The first term represents the damped free-streaming solution, while the second term encodes both the gradient expansion and non-perturbative corrections through the incomplete gamma function $\gamma(k, \xi_0)$. We note the similarity of the expression to the singulant formalism introduced in \cite{Heller:2021yjh}.   

This structure can be made more transparent by explicitly separating the perturbative and non-perturbative parts,
\begin{align}\label{Eq:MomFormSolInt}
    \rho_{n,l}(\tau) &= \sum_{k=0}^{\infty}   \lrrb{-\tau_R \mathbf{\hat{D}}  }^{k} \rho^{\text{eq}}_{n,l}(\tau) \nonumber\\
    &- e^{-\xi_0}e^{-\mathbf{\hat{K}}(\tau,\tau_0)}\lrb{ \sum_{k=0}^{\infty}   \lrrb{-\tau_R \mathbf{\hat{D}}  }^{k} \rho^{\text{eq}}_{n,l}(\tau_0)  -  \rho_{n,l}(\tau_0) }\,,
\end{align}

 Equation \eqref{Eq:MomFormSolTyl},\eqref{Eq:MomFormSolInt} provides a central result of this work. It shows explicitly that the exact solution can be decomposed into gradient expansion and non-perturbative contribution that carries information about initial conditions. Unlike standard approaches, this decomposition is obtained fully analytically and remains valid beyond the regime of small gradients. 

The system asymptotically approaches the gradient expansion, while the contribution from free-streaming modes encoded in the second term—decays exponentially. The expression in brackets in \eqref{Eq:MomFormSolInt} quantifies the discrepancy between the exact initial condition and its gradient expansion. However, it is important to note that neither of the two terms in \eqref{Eq:MomFormSolInt} is convergent by itself. This behavior mirrors the expansion obtained for the full distribution function, where the exact solution contains both a divergent asymptotic gradient series and a non-perturbative term carrying initial data.

\subsection{Dynamics of anisotropy}

To derive the hydrodynamic equations for the $0~+~1$D system, we begin by decomposing the moments into equilibrium and non-equilibrium ( anisotropic ) parts,
\begin{equation}
    \rho_{n,l} = \rho_{n,l}^{\rm eq} + \pi_{n,l}\,
\end{equation}
where $\rho_{n,l}^{\rm eq}$ given in \eqref{Eq:ZoEqlMom} is a function of $T$ and $\mu$. The evolution of $T$ and $\mu$ or alternatively $\varepsilon$ and $n$ is obtained from the conservation equation with the Landau matching constraint,
\begin{equation}\label{Eq:MomLandau}
    \begin{aligned}
    \rho_{0,0} &= \rho_{0,0}^{\rm eq}(T,\mu)\,,\\
    \rho_{1,0} &= \rho_{1,0}^{\rm eq}(T,\mu)\,.
 \end{aligned}
\end{equation}
This leads to the two conservation equations,
\begin{equation}\label{Eq:ZoMomCtn}
    \begin{aligned}
    \mathbf{\hat{D}}\rho_{0,0} &= 0 && -~ \text{(particle number conservation)}\,,\\
    \mathbf{\hat{D}}\rho_{1,0} &= 0 && -~ \text{(energy conservation)}\,,
 \end{aligned}
\end{equation}
which govern the dynamics of the equilibrium variables. The evolution of the non-equilibrium components $\pi_{n,l}$ follows directly from \eqref{Eq:ZoRTAMOmOp},
\begin{align}\label{Eq:ZoMomPiEvol}
     \partial_{\tau}\pi_{n,l} + \frac{\pi_{n,l} }{\tau_R} = - \mathbf{\hat{D}} \rho^{\rm eq}_{n,l}(\tau)-\mathbf{\hat{F}}\pi_{n,l}\,.
\end{align}
Together, equations \eqref{Eq:ZoMomCtn} and \eqref{Eq:ZoMomPiEvol} form a complete set describing the system’s evolution and are equivalent to the full moment hierarchy \eqref{Eq:ZoRTAMomeq}. We can obtain a Chapman-Enskog type gradient expansion for $\pi_{n,l}$ iterating on \eqref{Eq:ZoMomPiEvol}, to get the gradient expansion,
\begin{equation}
    \pi_{n,l}^{G} = \sum_{k=1}^{\infty}\lrrb{-\tau_R \mathbf{\hat{D}} }^{k} \rho^{\rm eq}_{n,l}(\tau)\,.
\end{equation}

Using the formal solution from \eqref{Eq:MomFormSol} or \eqref{Eq:MomFormSolInt}, we can express the solution to \eqref{Eq:ZoMomPiEvol} as,
\begin{align}\label{Eq:ZoPiSol}
    \pi_{n,l}   &=\sum_{k=1}^{\infty}   \lrrb{-\tau_R \mathbf{\hat{D}}  }^{k} \rho^{\text{eq}}_{n,l}(\tau) \nonumber\\
    &+ e^{-\xi_0}e^{-\mathbf{\hat{K}}(\tau,\tau_0)}\lrb{\rho_{n,l}(\tau_0) -  \sum_{k=0}^{\infty}   \lrrb{ -\tau_R \mathbf{\hat{D}}  }^{k} \rho^{\text{eq}}_{n,l}(\tau_0)   }\,.
\end{align}
This exact expression motivates us to subdivide the non-equilibrium components into gradient $\pi_{n,l}^{G}$ and the exponentially decaying non-perturbative $\pi_{n,l}^{T}$ terms ,
\begin{equation}
   \pi_{n,l} = \pi_{n,l}^{G}+\pi_{n,l}^{T} \,.
\end{equation}
The non-perturbative term being,
\begin{equation}
   \pi_{n,l}^{T}(\tau) = e^{-\xi_0}e^{-\mathbf{\hat{K}}(\tau,\tau_0)}\lrb{\rho_{n,l}(\tau_0) -  \sum_{k=0}^{\infty}   \lrrb{ -\tau_R \mathbf{\hat{D}}  }^{k} \rho^{\text{eq}}_{n,l}(\tau_0)   }\,.
\end{equation}
Note that the term enclosed in the brackets of \eqref{Eq:ZoPiSol} is $\pi_{n,l}^{T}(\tau_0) =  \pi_{n,l}(\tau_0) -\pi_{n,l}^{G}(\tau_0) $ with ${\pi}_{n,l}(\tau_0)  =  {\rho}_{n,l}(\tau_0) -  {\rho}_{n,l}^{~\text{eq}}(\tau_0)$, which can be thought of as a measure of how well the gradient expansion at $\tau_0$ captures the initial data. 

The corresponding evolution equations for the anisotropic components are,
\begin{align}
    \mathbf{\hat{D}}\pi_{n,l}^{G}  &= -\frac{\pi_{n,l}^{G} }{\tau_R}- \mathbf{\hat{D}} \rho^{\rm eq}_{n,l}\label{Eq:ZoMomPiGEvol}\\
    \mathbf{\hat{D}}\pi^{T}_{n,l}  &= - \frac{\pi^{T}_{n,l}}{\tau_R}\label{Eq:ZoPiTEvol}
\end{align}

A particularly striking observation is that the evolution equation \eqref{Eq:ZoMomPiGEvol} for the gradient component $\pi_{n,l}^{G}$ is structurally identical to the full non-equilibrium evolution equation \eqref{Eq:ZoMomPiEvol}. Moreover, the non-perturbative term can be expressed through the gradient contributions:
\begin{equation}\label{Eq:ZoPiTAltSol}
    \pi_{n,l}^{T}(\tau) \sim e^{-\xi_0}\sum_{k=0}^{\infty}\lrb{\sum_{l=0}^{k}\frac{\xi_0^k}{k!}}  \lrrb{ -\tau_R \mathbf{\hat{D}}  }^{k} \rho^{\text{eq}}_{n,l}(\tau)
\end{equation}

These observations are central to explaining the effectiveness of hydrodynamics even far from local equilibrium. They indicate that one can construct hydrodynamics starting from the gradient expansion, provided the transport coefficients are suitably adjusted. This approach yields evolution equations that reproduce the exact dynamics, and it closely mirrors the heuristic used in standard derivations of relativistic hydrodynamics. We demonstrate this explicitly by deriving the exact hydrodynamic equations in the following subsection and approximating them with rescaled transport coefficients.

\subsection{Hydrodynamic Equations}

Building on the insights from the previous section, we now derive the hydrodynamic evolution equations from the anisotropic moment equations. In the $0~+~1$D setting, the relevant quantities are the energy density $\varepsilon$, the bulk viscous pressure $\Pi$ and the shear pressure $\pi$. Our goal is to obtain a closed set of dynamical equations for these quantities.

Energy-momentum conservation provides the evolution equation for the energy density,
\begin{equation}
    \partial_{\tau}\varepsilon = -\frac{\varepsilon + P_L}{\tau}\,,
\end{equation}
where the longitudinal pressure $P_L = P + \Pi-\pi$. 
The pressure anisotropies are related to the anisotropic moments $\pi_{1,1} = \Pi -\pi$ and $\pi_{-1,0} = -\frac{3}{m^2}\Pi $. From the equations for the anisotropic components we get,
\begin{align}
  \partial_{\tau}\pi + \frac{\pi }{\tau_R} &= - \frac{8}{3\tau}\pi + \frac{5}{3\tau}\Pi - \frac{\beta_{\pi}}{\tau}  \nonumber\\
  & -\frac{m^2}{3}\frac{\pi_{-1,1} }{\tau}-\frac{\pi_{1,2}}{\tau} \label{Eq:ZoHShrExt}\\
  \partial_{\tau} \Pi+ \frac{\Pi }{\tau_R}&= -\lrrb{\frac{4}{3} - c_{s}^{2}}\frac{\Pi}{\tau} + \lrrb{\frac{1}{3} - c_{s}^{2}}\pi- \frac{\beta_{\Pi}}{\tau} \nonumber\\
  &- \frac{m^2}{3} \frac{1}{\tau}\pi_{-1,1}\label{Eq:ZoHBlkExt}\,.
\end{align}
The detailed derivation and the transport coefficients $\beta_{\pi}$ and $\beta_{\Pi}$ can be found in appendix \ref{Ap:ZpODHyd}. The evolution equations \eqref{Eq:ZoHBlkExt} and \eqref{Eq:ZoHShrExt} closely resemble those found in second-order hydrodynamics derived from traditional gradient expansion techniques \cite{Denicol:2010xn,Denicol:2012cn,Jaiswal:2013npa,Jaiswal:2013vta,Panda:2020zhr,Panda:2021pvq}, prior to the implementation of moment closure. To close the system of equations, the higher order moment couplings, $\pi_{-1,1}$ and $\pi_{1,2}$, must be expressed in terms of the hydrodynamic variables $\varepsilon$,$\text{n}$, $\Pi$, and $\pi$. This can be achieved, in principle, by utilizing the expressions derived in equation \eqref{Eq:MomFormSolTyl},
\begin{align}\label{Eq:MomClosure}
    \pi_{-1,1} &= e^{-\xi_0}\lrb{e^{-\mathbf{\hat{K}}(\tau,\tau_0)} \rho_{-1,1}(\tau_0) -\rho_{-1,1}^{\text{eq}}(\tau)}\nonumber\\
   &+  \sum_{k=1}^{\infty} \frac{ \gamma(k+1,\xi_0)}{k!} \lrrb{-\tau_R \mathbf{\hat{D}}  }^{k} \rho^{\text{eq}}_{-1,1}(\tau)\,,\\
   \pi_{~1,2} &= e^{-\xi_0} \lrb{e^{-\mathbf{\hat{K}}(\tau,\tau_0)} \rho_{1,2}(\tau_0) -\rho_{1,2}^{\text{eq}}(\tau)}\nonumber\\
   &+  \sum_{k=1}^{\infty} \frac{ \gamma(k+1,\xi_0)}{k!} \lrrb{-\tau_R \mathbf{\hat{D}}  }^{k} \rho^{\text{eq}}_{1,2}(\tau)\,.
\end{align}

% However we see that a complete closure is not possible as the free streaming initial conditions $e^{-\xi_0}e^{-\mathbf{\hat{K}}(\tau,\tau_0)} \rho_{-1,1}(\tau_0)$ cannot be captured solely by gradient terms. Nonetheless, if
% \begin{equation}\label{Eq:ZoHydVal}
%     \frac{e^{-\xi_0}\lrb{e^{-\mathbf{\hat{K}}(\tau,\tau_0)} \rho_{n,l}(\tau_0) -  \rho^{\text{eq}}_{n,l}(\tau)}}{\pi_{n,l}} \ll 1
% \end{equation}
% for $(n,l) = (-1,1),~(1,2)$, we can approximately close the system and write,
% \begin{align}
% \partial_{\tau} \Pi &+ \frac{\Pi }{\tau_R} 
%     = \frac{\Pi}{\tau}-   \mathbf{\hat{D}} P \nonumber\\
%                &+ \frac{m^2}{3}\frac{1}{\tau}\lrrb{ \sum_{k=1}^{\infty} \frac{ \gamma(k+1,\xi_0)}{k!} \lrrb{-\tau_R \mathbf{\hat{D}}  }^{k} \rho^{\text{eq}}_{-1,1}(\tau)}\label{Eq:ZoHBlkCls}\,,\\
% \partial_{\tau}\pi &+ \frac{\pi  }{\tau_R}  = - (4-c_{s}^{2})\frac{\pi}{\tau} + 2(2-c_{s}^2)\frac{\Pi}{\tau}\nonumber\\
%     &- \frac{1}{\tau}\lrrb{ \sum_{k=1}^{\infty} \frac{ \gamma(k+1,\xi_0)}{k!} \lrrb{-\tau_R \mathbf{\hat{D}}  }^{k} \rho^{\text{eq}}_{1,2}(\tau) } \nonumber\\
%     &- \frac{m^2}{3}\frac{1}{\tau} \lrrb{ \sum_{k=1}^{\infty} \frac{ \gamma(k+1,\xi_0)}{k!} \lrrb{-\tau_R \mathbf{\hat{D}}  }^{k} \rho^{\text{eq}}_{-1,1}(\tau)}\label{Eq:ZoHShrCls}\,.
% \end{align}

These expressions coincide with those obtained from the gradient expansion, with the crucial difference that each gradient term ( and therefore the transport coefficients) are rescaled by non-perturbative corrections arising from the $\gamma(k+1,\xi_0)$ terms. The pure gradient terms in the expansion gives near equilibrium, long time (long wavelength) dynamics while the exponentially decaying corrections contains memory effects of early time (short wavelength) free streaming effects. 

The gradient corrections due to the higher order moments enter the hydrodynamic evolution equations \eqref{Eq:ZoHBlkExt} and \eqref{Eq:ZoHShrExt} is of the order of $\mathcal{O}(1/\tau^k)$, $k \geq 2$ and follow the usual gradient expanded hydrodynamic structure with each order scaled by $\gamma(k+1,\xi_0)$ factor. These are expressible in terms of the hydrodynamic variables $\varepsilon$, $n$, $\Pi$, and $\pi$ (see Appendix \eqref{Ap:ZpODHyd}). However the first term which includes the free streaming corrections expressed in equations \eqref{Eq:MomClosure} is of the order of $\mathcal{O}(1/\tau)$. We can incorporate these terms into the transport coefficient $\beta_{\pi/\Pi}$. 
\begin{align}
\beta_{\pi} &\to \beta_{\pi} ~-(1/3)m^2\pi_{-1,1}^{(0)}-\pi_{1,2}^{(0)}\,,\\
    \beta_{\Pi} &\to \beta_{\Pi} -(1/3)m^2\pi_{-1,1}^{(0)}\,.
\end{align}

We can get from \eqref{Eq:FSExSol}
\begin{align}\label{Eq:ZeroMomClo}
    \pi_{-1,1}^{(0)} &\approx e^{-\xi_0}\lrrb{~\sum_{k=0}^{\infty}\, K_{-1,1,k}(\tau,\tau_{0})\rho_{-1,1+k}(\tau_0) - \rho^{\rm eq}_{-1,1}(\tau)}\,,\\
    \pi_{1,2}^{(0)}&\approx e^{-\xi_0}\lrrb{~\sum_{k=0}^{\infty}\, K_{~1,2,k}(\tau,\tau_{0})\rho_{1,2+k}(\tau_0) ~- \rho^{\rm eq}_{~1,2}(\tau)}\,,
\end{align}
where the initial moments $\rho_{n,l+k}(\tau_0)$ which contain the initial moment distribution, can be model parameters that can be obtained by fitting to data. 

The gradient terms can in general be written in the form,
\begin{align}
     \sum_{k=1}^{\infty} \frac{ \gamma(k+1,\xi_0)}{k!} \lrrb{-\tau_R \mathbf{\hat{D}}  }^{k} \rho^{\text{eq}}_{n,l}(\tau) &= {\rm a}_{n,l}~ \epsilon + {\rm b}_{n,l}~{\rm n} \nonumber\\
     & + {\rm c}_{n,l}~\Pi + {\rm d}_{n,l}~\pi
\end{align}
where ${\rm a}_{n,l},~{\rm b}_{n,l},~{\rm c}_{n,l},~{\rm d}_{n,l}$ are functions of the hydrodynamic variables $\epsilon,~n~,\Pi,\pi$. This will effectively close the equations \eqref{Eq:ZoHBlkExt}, \eqref{Eq:ZoHShrExt} and absorb the higher order contributions into the transport coefficients.

%%{\color{Red}A graph here would be nice}

\section{Generalisation to $3~+~1$D }

We now generalize the key insights obtained in the $0~+~1$D systems to the full $3~+~1$D case. In particular, we show that the structural equivalence between the gradient and non-perturbative contribution persists in higher dimensions, and that the exact dynamics can be reproduced through appropriately modified transport coefficients. 

The extension of our argument to the full $3~+~1$D case rests on two key observations drawn from the $0~+~1$D system. First, the evolution equations governing the exact non-equilibrium components structurally resemble those governing the gradient expansion terms. Second, the non-perturbative, exponentially decaying corrections that carry initial data can be written in terms of scaled versions of the local gradients. In this section, we demonstrate how these properties persist in the $3~+~1$D framework and explore their implications for formulating hydrodynamic equations. 

\subsection{The linearised moment equations}

Motivated by the moment evolution equations of the Boltzmann equation \cite{Denicol:2012cn,Ambrus:2023qcl} , we consider a generalized relaxation-type model,
\begin{equation}\label{Eq:ThpOMomEvol}
    \lrrb{\partial_{\tau} + \hat{\textbf{F}}}\vec{\rho} = -\hat{\textbf{C}}\lrrb{\vec{\rho} - \vec{\rho}^{~\text{eq}}} \,,
\end{equation}
with $ \partial_{\tau}  \equiv u^{\mu}\partial_{\mu}$. We note that the structure of the linearized moment equations parallels that of the Boltzmann equation itself. This equation can be thought of as a moment version of the RTA Boltzmann equation with $\hat{{\mathbf{C}}}[\vec{\rho}] \equiv \hat{\boldsymbol{\nu}}_{R}\lrrb{\vec{\rho} - \vec{\rho}^{~\text{eq}}}$ being a linearization of the Boltzmann collision kernel, typically around equilibrium.  This has the added advantage of being able to incorporate multiple time scales and therefore more generalized collision kernels. This structure typically emerges in momentum dependent RTA models \cite{Rocha:2021zcw, Biswas:2022cla, Gangadharan:2025qbp}.

The spectrum of $\hat{\mathbf{C}}$ should be positive semi-definite much like that of the spectrum of the linearized Boltzmann collision kernel. Since $\hat{\mathbf{C}}$ acts on an infinite-dimensional vector space, its spectrum typically contains a continuous component.  The kernel of $\hat{\mathbf{C}}$ comprises the five conserved moments associated with energy, momentum, and number conservation. The operator $\hat{\boldsymbol{C}}$ is positive semi-definite, as it corresponds to the collision frequency term projected into the moment space—generally non-zero.  In the Anderson–Witting form of the RTA, $\hat{\boldsymbol{\nu}}_{R}$ reduces to $\hat{\mathbf{I}}/\tau_R$ with a single eigenmode and decay rate given by the relaxation time $\tau_R$ and the conservation laws being imposed via matchig conditions. 

Here $\mathbf{\hat{F}}$ is the moment space Liouville operator. The operator $\mathbf{\hat{F}}$ has a the structure \cite{ Rocha:2021lze, deBrito:2023tgb, deBrito:2024qow},
\begin{align}
  \mathbf{\hat{F}}  = \mathbf{F}_{\mu}\nabla^{\mu} + \mathbf{A}\,
\end{align}
where $\nabla^{\mu}$ are the commoving derivatives and $\mathbf{F}_{\mu}$, $\mathbf{A}$ are matrix(linear) operators that couples the moments. Both $\partial_{\mu}$ and $\mathbf{\hat{F}}$ are dependent on the choice of frame i.e the definition of $u^{\mu}$.

\subsection{Solution in $3~+~1$ D}
For simplicity, we limit our analysis to the RTA collision kernel, where $\hat{\boldsymbol{\nu}}_{R} = \hat{\mathbf{I}} / \tau_R$. Most of the discussion in this section does not rely on this specific form of $\hat{\boldsymbol{\nu}}_{R}$; it only requires that the operator be invertible with positive definite eigenvalues. However, certain explicit results—such as the modifications to the transport coefficients—are derived under the assumption of the AW-RTA form. 

Under a general relaxation time model, the evolution equation for the moments is,

\begin{equation}\label{Eq:ThpOMomEvolRTA}
\mathbf{\hat{D}}\vec{\rho} = -\hat{\boldsymbol{\nu}}_{R} \lrb{\vec{\rho}-\vec{\rho}^{~eq}}\,,
\end{equation}
where $\mathbf{\hat{D}} = \lrrb{\partial_{\tau} + \hat{\textbf{F}}}$. 

Decomposing the moments into equilibrium and non-equilibrium parts, $\vec{\rho} = \vec{\rho}^{~eq} + \vec{\pi}$, the evolution of the non-equilibrium contribution becomes
\begin{equation}\label{Eq:ThpOMomPiEvol}
    \mathbf{\hat{D}}\vec{\pi}(\tau)  = -\hat{\boldsymbol{\nu}}_{R}~\vec{\pi}- \mathbf{\hat{D}}\vec{\rho}^{~eq} \,.
\end{equation}
We further split $\vec{\pi}$ into a gradient expansion piece and a non-perturbative piece, $\vec{\pi} = \vec{\pi}^{G} + \vec{\pi}^{T}$. The iterative gradient expansion yields
\begin{equation}\label{Eq:ThpOPiGExp}
     \vec{\pi}^{G}(\tau) = \sum_{k=1}^{\infty} \lrrb{- \hat{\boldsymbol{\nu}}_{R}^{-1}\mathbf{\hat{D}}}^k \vec{\rho}^{~eq}(\tau)\,.
\end{equation}
As $ \vec{\pi}^{G}$ satisfies \eqref{Eq:ThpOMomPiEvol}, the non-perturbative terms should satisfy, 
\begin{equation}\label{Eq:ThpOPiTEvol}
     \mathbf{\hat{D}}{ \vec{\pi}^{T}} = - \hat{\boldsymbol{\nu}}_{R}~ \vec{\pi}^{T}\,.
\end{equation}
This shows that even in 3+1D the structural equivalence between the gradient expansion and the exact anisotropy evolution continues to hold. The next step is to demonstrate that the exact evolution can be reproduced by suitably modifying the transport coefficients. To this end, we construct a formal solution of the moment evolution equation.

The moment equation \eqref{Eq:ThpOMomEvolRTA} is a linear time-evolution equation structurally similar to the Schrödinger equation. A formal solution can therefore be constructed using standard perturbative methods familiar from quantum mechanics. The extension to this setting, however, introduces new subtleties. In particular, unlike in the $0~+~1$D case presented earlier, the operators $\hat{\textbf{F}}$ and $\hat{\boldsymbol{\nu}}_{R}$ generally do not commute, as reflected in the non-vanishing commutator $[\mathbf{F}_{\mu}\nabla^{\mu}, \hat{\boldsymbol{\nu}}_{R}] \neq 0$.

\subsubsection*{Solutions to the moment equation}
 Consider the free streaming equation,
\begin{align}
    \lrrb{\partial_{\tau} + \mathbf{\hat{F}}}\vec{\rho}  = 0,
\end{align}
for which we can write down a propagator which satisfies,
\begin{equation}
    \partial_{\tau}\hat{\mathcal{F}}_{\tau,\tau_0} = -\mathbf{\hat{F}} \hat{\mathcal{F}}_{\tau,\tau_0}\,.
\end{equation}
The free streaming solution can be written as,
\begin{equation}
    \vec{\rho}^{~F}(\tau) = \hat{\mathcal{F}}_{\tau,\tau_0}\vec{\rho}(\tau_0) 
\end{equation}

Now consider the damped equation,
\begin{align}\label{Eq:ThpOMomDmp}
    \lrrb{\partial_{\tau} + \mathbf{\hat{F}}+ \hat{\boldsymbol{\nu}}_{R} }\vec{\rho}  = 0\,.
\end{align}

To solve \eqref{Eq:ThpOMomDmp}, we treat the damping term $\hat{\mathbf{\nu}}_{R}$ as an “interaction,” in analogy with the interaction (or collision) picture. We introduce the collision–picture moments
\begin{equation}
    \vec{\rho}_{c}(\tau) = \hat{\mathcal{F}}^{-1}_{\tau,\tau_0}\vec{\rho}(\tau)\,,
\end{equation}
together with the transformed operator
\begin{equation}
    \hat{\mathbf{\nu}}_{c}(\tau) = \hat{\mathcal{F}}^{-1}_{\tau,\tau_0}~\hat{\boldsymbol{\nu}}_{R}(\tau) ~\hat{\mathcal{F}}_{\tau,\tau_0}\,.
\end{equation}

In this representation, the evolution equation becomes,
\begin{equation}
    \lrrb{\partial_{\tau} + \hat{\boldsymbol{\nu}}_{c}}~\vec{\rho}_{c}(\tau) = \hat{\boldsymbol{\nu}}_{c} ~\vec{\rho}^{~\text{eq}}_{c}(\tau)
\end{equation}

Its solution is given by
\begin{equation}\label{Eq:ThpOMomDmpSol}
    \vec{\rho}_{c}{(\tau)} = \hat{\mathcal{W}}_{\tau,\tau_0} ~\vec{\rho}_{c}{(\tau_0)} + \int_{\tau_0}^{\tau} d\tau'\hat{\mathcal{W}}_{\tau,\tau'}~\hat{\boldsymbol{\nu}}_{c} ~\vec{\rho}_{c}^{~\text{eq}}(\tau')\,,
\end{equation}
where $ \hat{\mathcal{W}}_{\tau,\tau'} = \mathcal{T}e^{-\int_{\tau'}^{\tau}d\tau''\hat{\boldsymbol{\nu}}_{c} }$ denotes the time–ordered exponential propagator satisfying,
\begin{equation}\label{Eq:ThpOColProp}
    \partial_{\tau}\hat{\mathcal{W}}_{\tau,\tau'} =- \hat{\boldsymbol{\nu}}_{c}~\hat{\mathcal{W}}_{\tau,\tau'}\,.
\end{equation}

We now consider the case where $[\mathbf{\hat{F}}, \hat{\boldsymbol{\nu}}_{R}] = 0$
holds at all times. In this situation the dynamics simplify, closely resembling the familiar $0~+~1$D case. In particular, we have $\hat{\boldsymbol{\nu}}_{c}=\hat{\boldsymbol{\nu}}_{R}$, and the propagator reduces to
\begin{equation}
    \hat{\mathcal{W}}_{\tau,\tau'} = e^{-\xi'}
\end{equation}
, with $\xi' = \int_{\tau'}^{\tau}\frac{d\tau''}{\tau_R}$. The solution in the Schrödinger picture then takes the form
\begin{align}
    \vec{\rho}(\tau) = e^{-\xi_0}\hat{\mathcal{F}}_{\tau,\tau_0}\vec{\rho}(\tau_0) + \int_{\tau_0}^{\tau} d\tau’ \, \frac{e^{-\xi’}}{\tau_R} \, \hat{\mathcal{F}}_{\tau,\tau’} \, \vec{\rho}^{~\text{eq}}(\tau’).
\end{align}

From this we can obtain the series expansion,
\begin{align}\label{Eq:MomThpOSolComm}
\vec{\rho}(\tau) &= e^{-\xi_0}\hat{\mathcal{F}}_{\tau,\tau_0}\vec{\rho}(\tau_0) \nonumber\\
&\quad + \sum_{k=0}^{\infty} \frac{\gamma(k+1,\xi_0)}{k!}\lrrb{-\tau_R \mathbf{\hat{D}}}^{k}\vec{\rho}^{~\text{eq}}(\tau).
\end{align}

Thus, the analysis of the commuting case proceeds in direct analogy with the $0~+~1$D result. The solution \eqref{Eq:MomThpOSolComm} in the commuting case is obtained by performing the coordinate transformation $[\tau’,\tau] \to [\xi’,0]$ and expanding the integrand functions in a Taylor series in $\xi’$. This procedure relies on $\xi’$ being monotonic and commuting with $\hat{\mathcal{F}}$, allowing it to act as a scalar parameter in $\hat{\mathcal{F}}_{0,\xi’}$. In the non-commuting case, such a simplification is no longer possible, and the modification of transport coefficients cannot be expressed in the same compact form. However we can still get expressions for the gradient and non-perturbative decompositions. 

Using the hydrodynamic generator in \eqref{Eq:ThpOMomDmpSol} together with the propagator property \eqref{Eq:ThpOColProp}, we have
\begin{align}
     \int_{\tau_0}^{\tau} d\hat{\mathcal{W}}_{\tau,\tau'}~\vec{\rho}_{c}^{~\text{eq}}(\tau') &= \hat{\mathcal{W}}_{\tau,\tau'}~\vec{\rho}_{c}^{~\text{eq}}(\tau') \Bigg|_{\tau_0}^{\tau}\nonumber\\
     &-\int_{\tau_0}^{\tau}d\hat{\mathcal{W}}_{\tau,\tau'}       \hat{\boldsymbol{\nu}}^{-1}_{c}\partial_{\tau'}\vec{\rho}_{c}^{~\text{eq}}(\tau')\,
\end{align}
where we have assumed $\hat{\boldsymbol{\nu}}$ is invertible. This can always be done by projecting the moment equations to the subspace orthogonal to the kernel of $\hat{\boldsymbol{C}}$.  

From this it follows that (Appendix \eqref{Ap:ThpOColP}),
\begin{align}\label{Eq:MomThpOSolNComm}
    \vec{\rho}(\tau) &= \sum_{k=0}^{\infty}\lrrb{-\hat{\boldsymbol{\nu}}_{R}^{-1}   \mathbf{\hat{D}}}^k\vec{\rho}^{~\text{eq}}(\tau)   \nonumber\\
    &+\hat{\mathcal{F}}_{\tau,\tau_0}\hat{\mathcal{W}}_{\tau,\tau_0} ~\lrb{\vec{\rho}{(\tau_0)} - \sum_{k=0}^{\infty}~ \lrrb{-\hat{\boldsymbol{\nu}}^{-1}_{R} \mathbf{\hat{D}} }^k\vec{\rho}^{~\text{eq}}(\tau_0)}\,.
\end{align}

Equation \eqref{Eq:MomThpOSolNComm} constitute the central result of $3~+~1$D analysis. It demonstrates that the full solution can be decomposed into gradient expansion and non-perturbative contribution , mirroring the structure found in the $0~+~1$D case. This confirms that the mechanism underlying the effectiveness of hydrodynamics is not restricted to highly symmetric systems.  

We can now write down an explicit expression for the non-perturbative component, 
\begin{equation}\label{Eq:MomThpOSolTAni}
   \vec{\pi}^{T}(\tau) =  \hat{\mathcal{F}}_{\tau,\tau_0}\hat{\mathcal{W}}_{\tau,\tau_0} \lrb{\vec{\pi}(\tau_0) -\vec{\pi}^{G}(\tau_0)}\,,
\end{equation}
where $\vec{\pi}(\tau_0)  =  \vec{\rho}(\tau_0) -  \vec{\rho}^{~\text{eq}}(\tau_0)$ and the term in the bracket is the initial non-perturbative anisotropy $\vec{\pi}^{T}(\tau_0) = \vec{\pi}(\tau_0) -\vec{\pi}^{G}(\tau_0)$ mirroring the $0~+~1$D expression.
 
For a solution that is analytic in $\tau$, each gradient term, $ \lrrb{-\hat{\boldsymbol{\nu}}^{-1}_{R}   \mathbf{\hat{D}}}^k\vec{\rho}^{~\text{eq}}(\tau_0)$ in  $\vec{\pi}^{G}(\tau_0)$ can be written in terms of the corresponding terms $ \lrrb{-\hat{\boldsymbol{\nu}}^{-1}_{R}   \mathbf{\hat{D}}}^k\vec{\rho}^{~\text{eq}}(\tau)$  via a Taylor expansion about $\tau$. The Taylor coefficients then gives the appropriate modifications to the gradient expansion. Deriving an explicit form analogous to the $0~+~1$D solution \eqref{Eq:ZoPiTAltSol} or the commuting case \eqref{Eq:MomThpOSolComm} is, however, algebraically involved. In Appendix \eqref{Ap:ThpOFSP}, we present these expressions explicitly for the Anderson–Witting kernel under suitable simplifying assumptions.

\subsection{Relativistic Hydrodynamics}\label{SS:RelHydReTc}
In this section, we provide a heuristic derivation for hydrodynamic equations in the $3~+~1$D setting from the moment equations and how the rescaled transport coefficients emerge. Consider the hierarchy of moment equations

\begin{equation}\label{Eq:ThpOMomEvolGen}
\partial_{\tau}\vec{\rho} = -\lrrb{ \hat{\boldsymbol{\nu}}_{R}+ \mathbf{\hat{F}}  } \vec{\rho} + \hat{\boldsymbol{\nu}}_{R}\vec{\rho}^{~eq}\,,
\end{equation}

Let $\mathcal{M}$ be the vector space of all moments. We define the 'hydrodynamic' subspace $\mathcal{H}_{\rm N}$ as a $\rm N$-dimensional subspace of $\mathcal{M}$ that includes, but is not restricted to, the moments, $T^{\mu\nu}, ~{\rm n^{\mu}}$. We assume that the operator $\hat{\boldsymbol{\nu}}_{R}$ depend only on the chosen hydrodynamic moments, while the equilibrium moment $\vec{\rho}^{~eq}$ depends on the equilibrium variables $u^{\mu}, T,\mu$.

Let $\mathbf{\hat{P}}_{\mathcal{H}}$ be the projection operator that projects to $\mathcal{H}$.  For $\vec{\rho} \in \mathcal{M}$, the hydrodynamic moments $\vec{\rho}_{\mathcal{H}}= \mathbf{\hat{P}}_{\mathcal{H}}\vec{\rho} ~\in~ \mathcal{H}_{\rm N}$ and non-hydrodynamic moments $\vec{\rho}_{\perp} = \lrrb{ \mathbf{\hat{I}} -\mathbf{\hat{P}}_{\mathcal{H}} }\vec{\rho} ~\in ~\mathcal{H}^{\perp}$ form a complimentary subspaces in $\mathcal{M} = \mathcal{H}_{\rm N} \oplus\mathcal{H}^{\perp}$. We rewrite these equations for the anisotropies $\vec{\pi} $ and project the evolution equation Eq.\eqref{Eq:ThpOMomEvolGen} into the hydrodynamic subspace, we get the hydrodynamic equations,

\begin{align}\label{Eq:ThpOMomHydEvol}
\partial_{\tau}\vec{\pi}_{\mathcal{H}} &= -\mathbf{\hat{P}}_{\mathcal{H}}\lrrb{ \hat{\boldsymbol{\nu}}_{R}+ \mathbf{\hat{F}}  }\vec{\pi}_{\mathcal{H}}  \nonumber\\
&~~~~-\mathbf{\hat{P}}_{\mathcal{H}}\lrrb{ \hat{\boldsymbol{\nu}}_{R}+ \mathbf{\hat{F}}  } \vec{\pi}_{\perp} -\mathbf{\hat{P}}_{\mathcal{H}}~\hat{\boldsymbol{\rm D}} \vec{\rho}^{~eq} \,.
\end{align}
Here, we assumed the projector is independent of spacetime. This can always be arranged with an appropriate choice of moment basis. To close the system of equations within the hydrodynamic subspace, we can use the solution obtained in Eq.\eqref{Eq:MomThpOSolNComm} and project it to the non-hydrodynamic subspace,
\begin{align}
    \vec{\pi}_{\perp} &= \vec{\pi}^{~G}_{\perp}
    +\vec{\pi}^{~T}_{\perp}\,,
\end{align}
where $\vec{\pi}^{~G}_{\perp}$ is the projection of the gradient term and $\vec{\pi}^{~T}_{\perp}$ is the projection of the transient term. The gradient term $\vec{\pi}^{~G}_{\perp}$ can be written in term of the hydrodynamic moments as they are made up of equilibrium gradients, $\lrrb{-\hat{\boldsymbol{\nu}}^{-1} \mathbf{\hat{D}} }^k\vec{\rho}^{~\text{eq}}(\tau)$. Specifically, it is possible to write ,
\begin{equation}
     \vec{\pi}_{\perp} = \mathbf{\hat{\Delta}}\lrb{\vec{\rho}_{\mathcal{H}}}\vec{\pi}_{\mathcal{H}} \,.
\end{equation}
which gives the closure scheme. Note that this does not contradict vector space linear independence; rather, the dynamic is such that the coefficients in the non-hydrodynamic moment basis can be written as functions of the coefficients of the hydrodynamic moment basis (See appendix \ref{Ap:EqGradClos}).

The initial condition term $\vec{\pi}^{~T}_{\perp}$, depends both on the hydrodynamic terms via $\hat{\mathcal{F}}_{\tau,\tau_0}\hat{\mathcal{W}}_{\tau,\tau_0}$ ( Eq. \eqref{Eq:MomThpOSolTAni} ) and on initial distribution. We can write this contribution as
\begin{equation}
    \mathbf{\hat{P}}_{\mathcal{H}}\lrrb{ \hat{\boldsymbol{\nu}}_{R}+ \mathbf{\hat{F}}  }\vec{\pi}^{~T}_{\perp} = \vec{\beta}^{~\rm T}\lrb{\vec{\rho}_{\mathcal{H}}, \tau_0}\,.
\end{equation}
We can then write the closed set of hydrodynamic equations,
\begin{align}\label{Eq:ThpCloHydEvol}
\partial_{\tau}\vec{\pi}_{\mathcal{H}} &= -\mathbf{\hat{P}}_{\mathcal{H}}\lrrb{ \hat{\boldsymbol{\nu}}_{R}+ \mathbf{\hat{F}}}\lrb{ \mathbf{\hat{I}} + \mathbf{\hat{\Delta}} } \vec{\pi}_{\mathcal{H}}  \nonumber\\
&~~~~ -\mathbf{\hat{P}}_{\mathcal{H}}~\hat{\boldsymbol{\rm D}} \vec{\rho}^{~eq} -  \vec{\beta}^{~\rm T}\lrb{\vec{\rho}_{\mathcal{H}}, \tau_0}\,.
\end{align}
The corrections to the transport coefficients comes both from $\mathbf{\hat{\Delta}}$ and $\vec{\beta}^{~\rm T}$ terms.

\section{Discussion}

Different hydrodynamic models correspond to different choices of the hydrodynamic subspace $\mathcal{H}_{N}$ and a corresponding closure scheme. For ideal hydrodynamics, the hydrodynamic subspace contains only moments corresponding to $\varepsilon,~ n,~u^{\mu}$ with a five dimensional hydrodynamic subspace $\mathcal{H}_{5}$. Along with the equation of state, these form a closed set of equations where the assumption of local thermal equilibrium gives the closure scheme,  $\vec{\pi} = \vec{0}$.

Navier-Stokes theory employs the same hydrodynamic subspace \(\mathcal{H}_5\), but adopts a first-order gradient closure scheme in which higher-order moments are set to their first order asymptotic values (see Eq.~\eqref{Eq:HydFirst}). Higher-order theories such as the Burnett and super-Burnett formulations introduce additional gradient corrections to this closure, while remaining asymptotic in nature.
\begin{figure}[!h]
    \centering
    \includegraphics[width=1.0\linewidth]{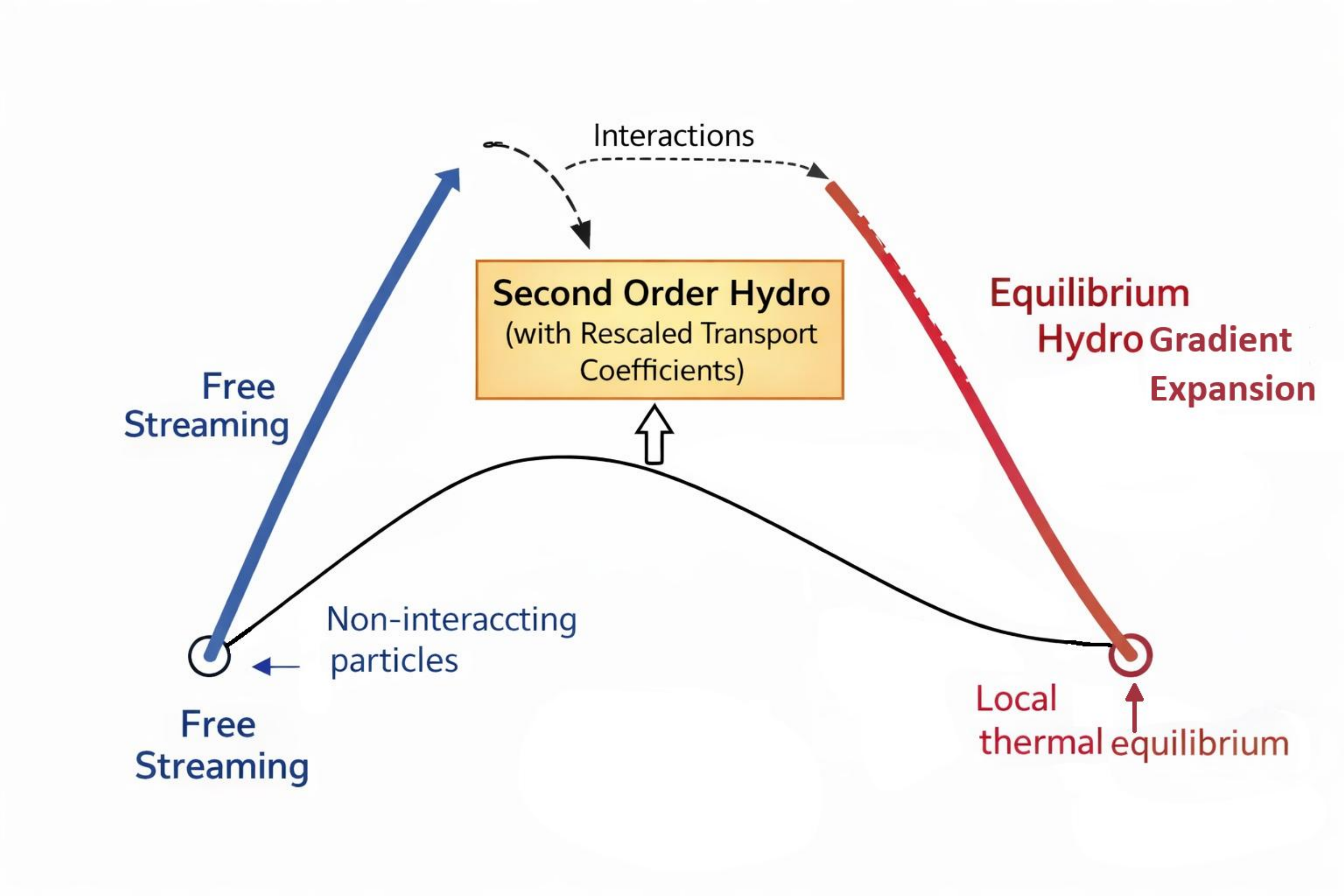}
    \caption{Relativistic hydrodynamic evolution $\sim$ gradient expansion with modified transport coefficients}
    \label{fig:FSfirHydro}
\end{figure}

In second-order Israel-Stewart type theories enlarge the hydrodynamic subspace is enlarged, introducing new variables $\pi^{\mu\nu}, ~\Pi,~ q^{\mu},~ V^{\mu}$ with their own dynamical equations. After a choice of frame, this adds an extra nine independent moments resulting in a fourteen-dimensional hydrodynamic subspace $\mathcal{H}_{19}$. The closure scheme again sets the higher-order moments to their asymptotic values. However, the new scheme is fundamentally different from the Navier-Stokes expansions in that the Israel-Stewart formulation incorporates non-asymptotic corrections. The resulting equations are not purely an expansion around equilibrium; in addition to introducing finite relaxation times for the anisotropic moments, they also allow for the description of free streaming dynamics. This is schematically illustrated in Fig. \ref{fig:FSfirHydro}. While Israel-Stewart theories were originally introduced to cure acausality and instability, their extended dynamical structure intrinsically goes beyond a purely asymptotic expansion around local equilibrium. 

Extending the hydrodynamic subspace to $N > 14$ with the asymptotic closure scheme for higher order moments allows a more accurate free streaming description. Near equilibrium, this will yield results similar to higher-order Chapman-Enskog type expansion. These two improvements address complementary regimes: enlarging the hydrodynamic subspace improves the description of far-from-equilibrium dynamics, while incorporating higher-order gradient corrections refines the near-equilibrium limit. A schematic of the subspace decomposition is given in \eqref{fig:hydro_hierarchy}.

\begin{figure*}[t]
\centering

\begin{tikzpicture}[
    font=\small,
    every node/.style={align=center},
    box/.style={draw, rounded corners, thick, inner sep=5pt, text width=4.2cm},
    arrow/.style={->, thick}
]

% ---------------------------
% Title
% ---------------------------
\node at (0,4.2) {\large Vector Space of Moments / Degrees of Freedom};

% ---------------------------
% Nested subspaces
% ---------------------------

\fill[gray!15] (0,0) ellipse (5.5 and 3.6);
\fill[blue!20] (0,0) ellipse (3.8 and 2.6);
\fill[green!25] (0,0) ellipse (2.6 and 1.3);

\draw[thick] (0,0) ellipse (5.5 and 3.6);

% Labels
\node at (0,1.9) {Israel--Stewart Subspace\\
$(e, n, u^\mu, \pi^{\mu\nu}, \Pi)$};

\node at (0,0) {Equilibrium Subspace\\
$(e, n, u^\mu)$};

\node at (0,-3.1) {Microscopic Degrees of Freedom};

% ---------------------------
% Theory boxes
% ---------------------------

% Left side
\node[box, fill=green!10] (eq) at (-7,1.5) {
\textbf{Equilibrium (Type 1)}\\
$\bullet$$e, n, u^\mu$\\
$\bullet$  Closure: Higher moments functions of $(T,\mu)$\\

};

\node[box, fill=green!20] (ns) at (-7,-1.5) {
\textbf{Navier--Stokes (Type 2)}\\
$\bullet$ $e, n, u^\mu$\\
$\bullet$ Gradient corrections\\
$\Pi^{\mu\nu} \sim \eta \partial^{\mu}u^{\nu}$\\
$\bullet$ $\eta(T,\mu) \sim$ Transport coefficients\\

};

% Right side
\node[box, fill=blue!10] (is) at (7,1.5) {
\textbf{Israel--Stewart (Type 3)}\\
$\bullet$ $e, n, u^\mu, \Pi^{\mu\nu}$\\
$\bullet$ $\Pi^{\mu\nu}$ independent\\
$\bullet$ $\eta(T,\mu) \sim$ Transport coefficients
};

\node[box, fill=blue!20] (ren) at (7,-1.5) {
\textbf{Renormalized (Type 4)}\\
$\bullet$ $e, n, u^\mu, \Pi^{\mu\nu}$\\
$\bullet$  Transport coefficients functions of $(T,\mu,\Pi^{\mu\nu},\dots)$\\
$\bullet$ Transport coefficients approximate\\ higher moment dynamics
};

% ---------------------------
% Arrows
% ---------------------------

\draw[arrow, shorten >=5pt, shorten <=5pt] (eq.east) -- (-1.5,0.8);
\draw[arrow, shorten >=5pt, shorten <=5pt] (ns.east) -- (-1.5,-0.8);

\draw[arrow, shorten >=5pt, shorten <=5pt] (2.5,1.2) -- (is.west);
\draw[arrow, shorten >=5pt, shorten <=5pt] (2.5,-1.2) -- (ren.west);
\draw[arrow, shorten >=5pt, shorten <=5pt] (1,-3) -- (ren.west);
\end{tikzpicture}

\caption{
Hierarchy of hydrodynamic descriptions as projections onto subspaces of the full moment space. Equilibrium and Navier--Stokes theories restrict to lower-dimensional vector space, while Israel--Stewart theories extend the set of independent variables, renormalized theories systematically include dynamical influence from ignored degrees of freedom.
}

\label{fig:hydro_hierarchy}

\end{figure*}

In section (\ref{SS:RelHydReTc}), we showed that the free streaming and relaxation type behaviour of higher order moments can be captured by altering the closure scheme to include transient contributions, rather than by introducing new dynamical variables. This provides a rigorous understanding of the renormalised transport coefficients proposed in previous studies \cite{Blaizot:2017ucy,Blaizot:2021cdv}. 

The results presented in this work suggest that the applicability of relativistic hydrodynamics is not fundamentally tied to proximity to local thermal equilibrium. Instead, hydrodynamics remains effective because it captures the dominant structure of the exact dynamics, with non-perturbative contributions modifying the transport coefficients in a systematic way. This provides a new theoretical basis for understanding hydrodynamiation in far from equilibrium system.

\section{Conclusion}

In this work, we have shown that solutions of the Boltzmann equation and its simplified models admit a universal structural decomposition into a Chapman--Enskog gradient contribution and an exponentially decaying non-perturbative mode. 
This result provides a systematic and explicit derivation of the trans-series structure underlying kinetic theory. 
We have further identified the origin of the divergence of the gradient expansion in terms of the coexistence of two fixed points of the collision dynamics, namely the collisionless regime and local thermal equilibrium, where the collision kernel vanishes. 
The gradient contribution captures the behaviour in the vicinity of equilibrium, while the non-perturbative terms encode the evolution towards the collisionless regime. 
A consistent description of collective dynamics in interacting systems therefore requires an interpolation between these two limits.

Israel--Stewart theory and its generalisations were originally introduced to address the causality issues inherent in the gradient expansion by promoting non-equilibrium contributions to independent dynamical variables with finite relaxation times. 
In this work, we have shown that these equations are not merely phenomenological extensions, but emerge naturally from the structure of the exact moment equations. 
In particular, we demonstrated that the evolution equations obtained from the gradient expansion are structurally identical to the exact dynamical equations governing momentum anisotropies. 
This provides a direct explanation for the effectiveness of hydrodynamics far from equilibrium. 
Furthermore, we have shown analytically that free-streaming behaviour can be reproduced within the gradient expansion through a systematic modification of transport coefficients, thereby providing a first-principles derivation of their renormalisation. 
In this way, our work both confirms and extends, through explicit analytical construction, the results reported in \cite{Blaizot:2017ucy,Blaizot:2019scw,Blaizot:2020gql,Blaizot:2021cdv,Jaiswal:2022udf}.

Relativistic hydrodynamics therefore emerges not as a strict expansion around local equilibrium, but as an effective theory that interpolates between free-streaming and collective dynamics. 
This perspective clarifies the role of non-hydrodynamic modes: rather than being discarded, their effects are systematically encoded into transport coefficients. 
In the context of quark--gluon plasma simulations, where transport coefficients are typically tuned to reproduce experimental observations, this provides a natural explanation for the remarkable success of hydrodynamics in describing far-from-equilibrium systems.

\textit{Future directions:}
Several extensions of the present framework would be of interest. 
First, it is important to generalise the analysis beyond the relaxation time approximation to more realistic collision kernels, where multiple relaxation scales may modify the structure of non-perturbative contributions and their impact on effective transport coefficients. Second, it would be desirable to formulate the present construction within a systematic projection operator framework, such as the Mori-Zwanzig formalism, applied directly at the level of the microscopic theory. Such an approach could provide a first-principles derivation of the hydrodynamic projection, and make explicit how non-hydrodynamic modes are encoded into renormalised transport coefficients. 
These directions may help establish a more general and unified connection between microscopic dynamics and emergent hydrodynamic behaviour.

\section{Acknowledgments}
 R.G. acknowledges support from the Department of Atomic Energy, Government of India. He thanks Victor Roy, Ankit Panda, and Krishanu Sengupta for insightful discussions, and Mahesh Gandikota for suggesting improvements to the draft manuscript.

\onecolumngrid

\appendix
\section{General Formulas}\label{Ap:GDef}

\subsection*{The Moment Integral}\label{Ap:MomInt}
 
In $0~+~1$D we define the moments as,
\begin{equation}
    \rho_{n,l} = \int \frac{d^3p}{(2\pi)^3} (p^{0})^{n}\lrb{\frac{p^{z}}{p^{0}\tau}}^{2l} f \,.
\end{equation}
For the equilibrium distribution function, we have
\begin{align}
    \rho_{n,l}^{\rm eq} &= \int \frac{d^3p}{(2\pi)^3} (p^{0})^{n}\lrb{\frac{p^{z}}{p^{0}}}^{2l} f^{\rm eq}\\
                    &= e^{\mu/T}\int \frac{d^3p}{(2\pi)^3}(p^{0})^{n}\lrb{\frac{p^z}{p^{0}}}^{2l} e^{-p^{0}/T}
\end{align}
The final integral is
 where $(p^{0})^2 = |p|^2 + m^2$. 
 Define,
 \begin{align}
     y &= \frac{p^{0}}{T}\,,\\
     z &= \frac{m}{T}
 \end{align}

 We have 
 \begin{equation}
     |p|^2 = T^2(y^2 - z^2 )
 \end{equation}
 So,
 \begin{equation}
     d|p| = T \frac{y}{\sqrt{y^2-z^2}}dy
 \end{equation}

 \begin{align}
    \rho_{n,l}^{\rm eq}(T,z) &= \frac{e^{\mu/T}}{(2\pi)^3}\int d|p|p^2\sin{\theta}d\theta d\phi  (p^{0})^{n-2l}|p|^{2l}\cos^{2l}{\theta}e^{-p^{0}/T} \nonumber\\
                    &= \frac{e^{\mu/T}}{2\pi^2}\frac{1}{2l + 1}\int T^2(y^2-z^2) dy \frac{Ty}{\sqrt{y^2-z^2}} \lrb{yT}^{n-2l}T^{2l}(y^2-z^2)^{l} e^{-y}\nonumber\\
                    &= \frac{e^{\mu/T}}{2\pi^2}\frac{T^{n+3}}{2l + 1} \int_{z}^{\infty} dy y^{n-2l+1}(y^2-z^2)^{l+1/2}e^{-y}
\end{align}
\begin{equation}
    \rho_{n,l}^{\rm eq}(T,z) =   \frac{T^{n+3}}{2\pi^2}\frac{G_{n + 3,l}(z) }{(2l + 1)}  e^{\mu/T} 
\end{equation}

We have 
\begin{align}
    \varepsilon  &= \rho^{\rm eq}_{1,0} = \frac{T^{4}}{2\pi^2}G_{4,l}(z)  e^{\mu/T}\\
    P  &= \rho^{\rm eq}_{1,1} = \frac{1}{3}\frac{T^{4}}{2\pi^2}G_{4,l}(z)  e^{\mu/T}
\end{align}
In general we have the relation,
\begin{align}\label{Eq:MOmEqinP}
     \rho^{\rm eq}_{1,l} &= \frac{1}{2l +1}\frac{T^{4}}{2\pi^2}G_{4,l}(z) e^{\mu/T} \\
     &= \frac{3}{(2l+1)} \frac{G_{4,l}(z)}{G_{4,1}(z)} P
\end{align}
and the conformal limit,
\begin{equation}
    \rho^{\rm eq}_{1,l} = \frac{3}{(2l+1)} P
\end{equation}

\subsubsection{The G-Function}
The function $G_{n,l}(z)$ appearing in the equilibrium moment is defined as,
\begin{align}
    G_{n,l}(z) &= \int_{z}^{\infty} dy y^{n-2l-2}(y^2-z^2)^{l+1/2}e^{-y}\\
\end {align}

We note the relation,
\begin{align} \label{EqAp:z2Gnl}
    z^2 G_{n,l}(z) &= \int_{z}^{\infty} dy z^2 y^{n-2l-2}(y^2-z^2)^{l+1/2}e^{-y} \\
                &= -\int_{z}^{\infty} dy \lrrb{(y^2 - z^2) - y^2} ~y^{n-2l-2}(y^2-z^2)^{l+1/2}e^{-y} \\
                &= -\int_{z}^{\infty} dy  y^{n-2l-2}(y^2-z^2)^{l+1+1/2}e^{-y} + \int_{z}^{\infty} dy  y^{n+2-2l-2}(y^2-z^2)^{l+1/2}e^{-y}\\
                &=G_{n+2,l}(z) - G_{n+2,l+1}(z)
\end{align}

For $z = 0$, we have
\begin{align}
    G_{n,l}(z = 0) &= \int_{z}^{\infty} dy y^{n-2l-2}(y^2-0^2)^{l+1/2}e^{-y}\nonumber\\
                        &= \int_{0}^{\infty} dy y^{n-1}e^{-y}\nonumber\\
                      &= \Gamma(n)
\end{align}

\section{$0~+~1$D Solution}\label{Ap:ZpOD}

%%\subsection{Exact solutions to the moment equation}\label{Ap:ZpODExt}
We rewrite the moment equations as
\begin{align}
     \lrrb{\partial_{\tau} + \lrb{\frac{2l+1}{\tau}\mathbf{\hat{I}}  - \frac{2l-n}{\tau}\mathbf{\hat{S}} } +  \frac{\mathbf{\hat{I}}}{\tau_{R}} }\rho_{n}(\tau,l)= \frac{\mathbf{\hat{I}}}{\tau_{R}}\rho_{n}^{\text{eq}} (\tau,l)
\end{align}
where we have defined the shift operator,
\begin{align}
    \mathbf{\hat{S}} \lrrb{h(l)} = h(l+1)\,.
\end{align}
We define the generator of free streaming, 
\begin{equation}
    \mathbf{\hat{F}} = \frac{2l+1}{\tau}\mathbf{\hat{I}}  - \frac{2l-n}{\tau}\mathbf{\hat{S}}\,.
\end{equation}
This essentially encodes the linear matrix algebra of the set of coupled differential equations. But, this allows us to use formal algebraic manipulations to obtain an integral form of the differential equations by considering them as first-order operator differential equations.

To find the propagator of the solutions, we define,
\begin{align}
   \hat {\mathcal{M}}_{n,l}= \int_{\tau_0}^{\tau}\lrb{\frac{2l+1}{\tau}\mathbf{\hat{I}}  - \frac{2l-n}{\tau}\mathbf{\hat{S}}  +  \frac{\mathbf{\hat{I}}}{\tau_{R}}}\,.
\end{align}
which gives us the integrating (propagator) factor,
\begin{align}
    \exp{\hat{\mathcal{M}}'_{n,l}} &= \exp{  \int_{\tau'}^{\tau}d\tau'\lrb{\frac{2l+1}{\tau'}\mathbf{\hat{I}}  - \frac{2l-n}{\tau'}\mathbf{\hat{S}}  +  +  \frac{\mathbf{\hat{I}}}{\tau_{R}} }}  
\end{align}
and its inverse
\begin{align}
    \exp{-\hat{\mathcal{M}}'_{n,l}} &= \exp{  -\int_{\tau'}^{\tau}\lrb{\frac{2l+1}{\tau} \mathbf{\hat{I}} - \frac{2l-n}{\tau}\mathbf{\hat{S}} }  +  +  \frac{\mathbf{\hat{I}}}{\tau_{R}} }  \,.
\end{align}
Such a manipulation is well defined as
\begin{align}
    [\hat{\mathcal{M}}_{n,l},\partial_{\tau}\hat{\mathcal{M}}_{n,l}] = 0 \,.
\end{align}

We now define the operator, which is the integral of the generator of Free streaming 
\begin{align}
    \mathbf{\hat{K}}_{\tau,\tau'} &= \int_{\tau'}^{\tau} \frac{2l+1}{\tau} \mathbf{\hat{I}}- \frac{2l-n}{\tau}\mathbf{\hat{S}} d\tau' \\
                        & = \log{\lrb{\frac{\tau}{\tau'}} } \lrrb{(2l+1)\mathbf{\hat{I}}-(2l-n) \mathbf{\hat{S}}_{l}}
\end{align}

This allows us to write a formal solution,
\begin{align}
    \rho_{n,l}(\tau) &= e^{-\xi_0}e^{ -\mathbf{\hat{K}}_{\tau,\tau_0} }\rho_{n,l}(\tau_0)
     + \int_{\tau_0}^{\tau}d\tau' \frac{1}{\tau_{R}}e^{-\xi'}e^{ -\mathbf{\hat{K}}_{\tau,\tau'}} \rho_{n,l}^{\text{eq}}(\tau')
\end{align}

To put this expression in a useful form, we note that if we define the operators,
\begin{align}
    \mathbf{\hat{X}} &= -\log{\lrb{\frac{\tau}{\tau'}}} \lrb{2l+1}\mathbf{\hat{I}}\\
    \mathbf{\hat{Y}}&= \log{\lrb{\frac{\tau}{\tau'}}}\lrb{2l-n}\mathbf{\hat{S}}\,,
\end{align}
They satisfy the commutation relation,
\begin{align}
    [\mathbf{\hat{X}},\mathbf{\hat{Y}}] = \alpha \mathbf{\hat{Y}} \,.
\end{align}
where $\alpha = -2\log{\lrb{\frac{\tau}{\tau'}}} $. For such operators, we have the property \cite{Suzuki:1985wzj},
\begin{equation}
    e^{\mathbf{\hat{X}}+ \mathbf{\hat{Y}}} = e^{\mathbf{\hat{X}}}e^{f(\alpha)\mathbf{\hat{Y}}}
\end{equation}
where 
\begin{equation}
    f(\alpha) = \frac{1 - e^{\alpha}}{\alpha} \,.
\end{equation}
Then we have
\begin{equation}
     f(\alpha) \mathbf{\hat{Y}}  = -\frac{(2l -n)}{2}\lrrb{1 - \lrb{\frac{\tau'}{\tau}}^{2}}\mathbf{\hat{S}}
\end{equation}

Using this we can write
\begin{align*}
    e^{ -\log{\lrb{\frac{\tau}{\tau'}} } \lrrb{(2l+1)\mathbf{\hat{I}}-(2l-n) \mathbf{\hat{S}}}} &=e^{ -\log{\lrb{\frac{\tau}{\tau'}} } \lrb{2l+1}\mathbf{\hat{I}}}e^{ -\frac{(2l -n)}{2}\lrb{1 - \lrb{\frac{\tau'}{\tau}}^{2}}\mathbf{\hat{S}} }\\
    &=\lrb{\frac{\tau'}{\tau}}^{2l+1} \sum_{k=0}^{\infty} \lrrb{-\frac{(2l -n)}{2}\lrb{1 - \lrb{\frac{\tau'}{\tau}}^{2}}\mathbf{\hat{S}}}^{k}\\
    &=\lrb{\frac{\tau'}{\tau}}^{2l+1}\sum_{k=0}^{\infty}\frac{(-1)^k(2l -n)^{(2k)}}{2^{k}k!} \lrrb{1 - \lrb{\frac{\tau'}{\tau}}^{2}}^{k}\mathbf{\hat{S}}^{k}
\end{align*}

where $(\cdot)^{(k)}$ is the rising Pochhammer symbol. 

We can immediately write down the free streaming solution,
\begin{align}\label{ApEq:FSExSol}
    \rho^{FS}_{n}(\tau,l) &=  \,\sum_{k=0}^{\infty}\frac{\lrb{l -\frac{n}{2}}^{(k)}}{k!} \lrb{\frac{\tau_0}{\tau}}^{2l+1}\lrrb{ \lrb{\frac{\tau_0}{\tau}}^{2} -1}^{k} \rho_{n}(\tau_0,l+k) 
\end{align}

We define the kernel,
\begin{equation}\label{ApEq:GenInt_Ker}
    K_{n,l,k}(\tau,\tau^{\prime}) =  \frac{\lrb{l -\frac{n}{2}}^{(k)}}{k!} \lrb{\frac{\tau^{\prime}}{\tau}}^{2l+1}\lrrb{1 - \lrb{\frac{\tau'}{\tau}}^{2}}^{k} \,,
\end{equation}
where $(\cdot)^(k)$ is the rising Poccammer symbol. Note that 
\begin{equation}
    K_{n,l,0}(\tau,\tau^{\prime}) = \lrb{\frac{\tau^{\prime}}{\tau}}^{2l+1} \,,
\end{equation}
and 
\begin{align}
    K_{n,l,k}(\tau,\tau) &= 1 && k = 0 \,,\\
    K_{n,l,k}(\tau,\tau) &= 0 && k \geq 1\,.
\end{align}

The solution to the relaxation equation can be written down as,
\begin{align}
    \rho_{n,l}(\tau) &= e^{-\xi_0}\sum_{k=0}^{\infty}\frac{\lrb{l -\frac{n}{2}}^{(k)}}{k!} \lrb{\frac{\tau_0}{\tau}}^{2l+1}\lrrb{ \lrb{\frac{\tau_0}{\tau}}^{2} -1}^{k} \rho_{n,l+k}(\tau_0) \nonumber \\
    & + \int_{\tau_0}^{\tau}d\tau' \frac{e^{-\xi'}}{\tau_{R}}\lrb{\frac{\tau'}{\tau}}^{2l+1}\sum_{k=0}^{\infty}\frac{(l -\frac{n}{2})^{(2k)}}{k!} \lrrb{ \lrb{\frac{\tau'}{\tau}}^{2}-1}^{k} \rho_{n,l+k}^{\text{eq}}(\tau)\,.
\end{align}

The generator of hydrodynamics is,
\begin{align}
    \rho^{G}_{n,l}(\tau) &=  \int_{\tau_0}^{\tau}d\tau' \frac{e^{-\xi'}}{\tau_{R}}\,\sum_{k=0}^{\infty} K_{n,l,k}(\tau,\tau^{\prime}) \rho_{n,l+k}^{\text{eq}}(\tau') 
\end{align}

We can now use the expression for $\rho_{n,l}^{\text{eq}}$,
\begin{align}
    \rho^{G}_{n,l}(\tau) &=  \int_{\tau_0}^{\tau}d\tau' \,\frac{e^{-\xi'}}{\tau_{R}}\,\sum_{k=0}^{\infty} K_{n,l,k}(\tau,\tau^{\prime}) \frac{T^{n+3}}{2\pi^2}\frac{G_{n+3,l+k}(z) }{(2(l+k) + 1)}  e^{\mu/T} 
\end{align}
For the conformal case  $z = 0$, we have $G_{n+3,l+k}(z) = \Gamma(n+3)$ which is independent of $l$ and $k$. This allows us the simplification,

\begin{align}
    \rho^{G}_{n,l}(\tau,z = 0) & =  \frac{\Gamma(n+3)}{(2\pi^2)(2l+1)}\int_{\tau_0}^{\tau}d\tau' \frac{e^{-\xi'}}{\tau_{R}}\lrb{\frac{\tau'}{\tau}}^{2l+1} T^{n+3}  \,\sum_{k=0}^{\infty}\frac{(1 + 1/2)^{(k)}\lrb{l -\frac{n}{2}}^{(k)}}{(l + 3/2)^{(k)}} \frac{1}{k!}  \lrrb{ \lrb{\frac{\tau'}{\tau}}^{2} -1}^{k} \\
    &= \frac{\Gamma(n+3)}{(2\pi^2)(2l+1)}\int_{\tau_0}^{\tau}d\tau' \frac{e^{-\xi'}}{\tau_{R}}\lrb{\frac{\tau'}{\tau}}^{2l+1}  F( l +\frac{1}{2},l -\frac{n}{2};l +\frac{3}{2};\lrb{\frac{\tau'}{\tau}}^{2} -1) \,T^{n+3}  \,,
\end{align}

where $F(a,b;c;x)$ is the hypergeometric function.

\subsection{$0~+~1$D Gradient expansion}\label{Ap:ZpODGrd}

Consider the hydrodynamic generator,
\begin{equation}
    \rho_{n,l}^{G}(\tau) =\int_{\tau_0}^{\tau}d\tau' \frac{1}{\tau_{R}}e^{-\xi'}e^{ -\mathbf{\hat{K}}_{\tau,\tau'}} \rho_{n,l}^{\text{eq}}(\tau')
\end{equation}
where $\xi' = \int_{\tau'}^{\tau}d\tau'\frac{1}{\tau_R}$. The free streaming propagator
\begin{equation}
    \mathcal{K}_{\tau,\tau'} = e^{ -\mathbf{\hat{K}}_{\tau,\tau'}}
\end{equation}
satisfies
\begin{equation}
    \partial_{\tau'}\mathcal{K}_{\tau,\tau'}\rho_{n,l}(\tau) = \mathcal{K}_{\tau,\tau'}\lrb{\partial_{\tau'} + \mathbf{\hat{F}} }\rho_{n,l}(\tau') 
\end{equation}

Consider the change of variables, $\tau' \to \xi' $. We now expand $e^{ -\mathbf{\hat{K}}(0,\xi')} \rho_{n,l}^{\text{eq}}(\xi')$ in  a Taylor series about $\xi' = 0$ as
\begin{align}
    e^{ -\hat{F}(0,\xi')} \rho_{n,l}^{\text{eq}}(\xi') &= \sum_{k = 0}^{\infty}\frac{(\xi')^k}{k!}\frac{\text{d}^k}{\text{d}\xi'^k}  e^{ -\mathbf{\hat{K}}(0,\xi')} \rho_{n,l}^{\text{eq}}(\xi')\Bigg |_{\xi' = 0}\\
    &= \sum_{k = 0}^{\infty}\frac{(\xi')^k}{k!}\lrrb{-\tau_R (\partial_{\tau'} + \mathbf{\hat{F}})}^{k}~\rho_{n,l}^{\text{eq}}(\xi')\Bigg |_{\tau' = \tau}\,.
\end{align}
For the $\xi'$ integration we note,
\begin{equation}
   \int_{\xi_0}^{0}d\xi' e^{-\xi'}(\xi')^k = \gamma(k+1,\xi_0)
\end{equation}

We now finally get the hydrodynamic generator in terms of the gradients,
\begin{equation}
     \rho_{n,l}^{G}(\tau) = \sum_{k = 0}^{\infty}\frac{\gamma(k+1,\xi_0)}{k!}\lrrb{-\tau_R (\partial_{\tau} + \mathbf{\hat{F}})}^{k}~\rho_{n,l}^{\text{eq}}(\tau) 
\end{equation}

\section{Hydrodynamic Equations}\label{Ap:ZpODHyd}

Hydrodynamics in $0~+~1$D requires us to find the dynamical equations for the energy density $\varepsilon$ and the two anisotropies, bulk pressure $\Pi$ and shear pressure $\pi$. From energy conservation we get the dynamical equation for energy density ($\varepsilon = \rho_{1,0} $),
\begin{align}
    \mathbf{\hat{D}}\rho_{0,0} = \mathbf{\hat{D}}\text{n} = 0 \\
    \mathbf{\hat{D}}\rho_{1,0} = \mathbf{\hat{D}}\varepsilon = 0  
\end{align}
which gives us the familiar energy density and number density evolution equation for Bjorken flow,
\begin{align}
     \partial_{\tau}\text{n}         &= - \frac{\text{n} }{\tau}\\
    \partial_{\tau}\varepsilon &= -\frac{\varepsilon + P_L}{\tau}
\end{align}
where $P_L$ is the longitudinal pressure $P_L = P + \Pi - \pi$. The equations for the non-equlibrium components,
\begin{equation}\label{Eq:ApNEqEvol}
     \mathbf{\hat{D}}\pi_{n,l}  = -\frac{\pi_{n,l} }{\tau_R}- \mathbf{\hat{D}} \rho^{\rm eq}_{n,l}
\end{equation}
will give us the evolution equations for the pressure anisotropies. To find this we note the relation between the anisotropic moments and the shear and bulk pressure. The bulk pressure can be computed from the moments using the relations,
\begin{align}
    \pi &= P + \Pi - P_L\\
    \Pi  &= \frac{1}{3}(\varepsilon - 3P) - \frac{1}{3}\langle m^2\rangle\,.
\end{align}

From this we can infer,
\begin{align}
    \Pi &= \frac{1}{3}(\varepsilon - 3P)- \frac{1}{3} m^2(\rho_{-1,0}^{\rm eq} + \pi_{-1,0})  \nonumber\\
        &=-\frac{m^2}{3}\pi_{-1,0}\\
    \pi &= \Pi-\pi_{1,1}\\
        &= -\frac{m^2}{3}\pi_{-1,0} - \pi_{1,1}
\end{align}

In the conformal limit
\begin{equation}
    \pi = - \pi_{1,1}
\end{equation}

\subsection{Speed of sound and bulk modulus}
Define 
\begin{align}\label{EqAp:GenSpBlk}
    c^2_{n,l}    &= \pdv{\rho^{\rm eq}_{n,l}}{\varepsilon}\,, \\ 
    \kappa_{n,l} &=n\pdv{\rho^{\rm eq}_{n,l}}{\text{n}}\,.
\end{align}
with the speed of sound $c_s^2 = c^2_{1,1}$ and bulk modulus $\kappa = \kappa_{1,1}$  as $\rho_{1,1}^{\rm eq} = P$, the equilibrium pressure. 

We now find explicit expressions for the variables defined in Eq\eqref{EqAp:GenSpBlk}. 

\begin{align}
\frac{\partial \rho_{n,l}^{eq}}{\partial T}
&= \frac{\partial}{\partial T}
\left[
e^{\alpha}\,\frac{T^{n+3}}{(2l+1)2\pi^{2}}\,G_{n+3,l}(z)
\right] \\[6pt]
&= e^{\alpha}\,
\frac{(n+3)T^{2}}{(2l+1)2\pi^{2}}\,G_{n+3,l}
\;+\;
e^{\alpha}\,
\frac{T^{3}}{(2l+1)2\pi^{2}}\,
\frac{\partial G_{n+3,l}}{\partial z}\,
\frac{\partial z}{\partial T}
\end{align}

Using
\[
\frac{\partial z}{\partial T} = -\frac{z}{T},
\qquad
\frac{\partial G_{n+3,l}}{\partial z} = \frac{1}{z}((n+3)G_{n+3,l}- G_{n+4,l}),
\]
we get
\begin{align}
    \frac{\partial G_{n+3,l}}{\partial T} = -\frac{1}{T}((n+3)G_{n+3,l}- G_{n+4,l})
\end{align}

\begin{align}
\frac{\partial \rho_{n,l}^{eq}}{\partial T}
&=
e^{\alpha}\,
\frac{(n+3)T^{n+2}}{(2l+1)2\pi^{2}}\,G_{n+3,l}- e^{\alpha}\,\frac{T^{n+3}}{(2l+1)2\pi^{2}}\,\frac{1}{T}\,((n+3)G_{n+3,l}- G_{n+4,l}) \\[6pt]
&=e^{\alpha}\,\frac{T^{n+2}}{(2l+1)2\pi^{2}}G_{n+4,l}
\end{align}

\begin{align}
\frac{\partial \rho_{n,l}^{eq}}{\partial \alpha}
&= \frac{\partial}{\partial \alpha}\left[e^{\alpha}\,\frac{T^{n+3}}{(2l+1)2\pi^{2}}\,G_{n+3,l}\right] \\[6pt]
&=e^{\alpha}\,\frac{T^{n+3}}{(2l+1)2\pi^{2}}\,G_{n+3,l}\;=\; \rho_{n,l}^{eq}
\end{align}

\begin{align}
    \frac{\partial \varepsilon}{\partial T} = e^{\alpha}\,\frac{T^{3}}{2\pi^{2}}G_{5,0} &&      \frac{\partial \varepsilon}{\partial \alpha} = e^{\alpha}\,\frac{T^{4}}{2\pi^{2}}G_{4,0}\\
    \frac{\partial \rm n}{\partial T} = e^{\alpha}\,\frac{T^{2}}{2\pi^{2}}G_{4,0} && \frac{\partial \rm n}{\partial \alpha} = e^{\alpha}\,\frac{T^3}{2\pi^{2}}G_{3,0}
\end{align}

We then have the Jacobian relation,
\begin{equation}\label{EqAp:EnTaJacob}
    \begin{bmatrix}
        {\rm d}\varepsilon\\
        {\rm d}\rm n
    \end{bmatrix} = e^{\alpha}\,\frac{T^{2}}{2\pi^{2}}\begin{bmatrix} TG_{5,0} && T^2 G_{4,0} \\
    G_{4,0} && T G_{3,0}   
    \end{bmatrix}\begin{bmatrix}
        {\rm d} T\\
        {\rm d} \alpha
    \end{bmatrix}
\end{equation}

The inverse Jacobian relation is 
\begin{equation}\label{EqAp:TaEnJacob}
    \begin{bmatrix}
        {\rm d} T\\
        {\rm d} \alpha
    \end{bmatrix} = e^{-\alpha}\,\frac{2\pi^{2}}{T^{4}}\frac{1}{G_{5,0}G_{3,0}-G_{4,0}^{2}}\begin{bmatrix} T G_{3,0}  && -T^2 G_{4,0} \\
    -G_{4,0} &&   TG_{5,0}
    \end{bmatrix}\begin{bmatrix}
        {\rm d}\varepsilon\\
        {\rm d}\rm n
    \end{bmatrix}
\end{equation}

From this we get the partial derivative relations

\begin{align}
    \frac{\partial T}{\partial \varepsilon} = ~~~e^{-\alpha}\,\frac{2\pi^{2}}{T^{3}}\frac{G_{3,0} }{G_{5,0}G_{3,0}-G_{4,0}^{2}}&&\frac{\partial T}{\partial \rm n}  =  -e^{-\alpha}\,\frac{2\pi^{2}}{T^{2}}\frac{G_{4,0} }{G_{5,0}G_{3,0}-G_{4,0}^{2}}   \\
    \frac{\partial \alpha}{\partial \varepsilon} = -e^{-\alpha}\,\frac{2\pi^{2}}{T^{4}}\frac{G_{4,0} }{G_{5,0}G_{3,0}-G_{4,0}^{2}} && \frac{\partial \alpha}{\partial \rm n} = ~~~~e^{-\alpha}\,\frac{2\pi^{2}}{T^{3}}\frac{G_{5,0} }{G_{5,0}G_{3,0}-G_{4,0}^{2}}
\end{align}

Now we can find
\begin{align}
     \frac{\partial \rho_{n,l}^{eq}}{\partial \varepsilon}  &=\frac{\partial \rho_{n,l}^{eq}}{\partial T} \frac{\partial T}{\partial \varepsilon} + \frac{\partial \rho_{n,l}^{eq}}{\partial \alpha} \frac{\partial \alpha}{\partial \varepsilon}\\
    &= \lrrb{e^{\alpha}\,\frac{T^{n+2}}{(2l+1)2\pi^{2}}G_{n+4,l}}\lrb{e^{-\alpha}\,\frac{2\pi^{2}}{T^{3}}\frac{G_{3,0} }{G_{5,0}G_{3,0}-G_{4,0}^{2}}} \\
    &~~~~~~~~~~~~+ \lrrb{e^{\alpha}\,\frac{T^{n+3}}{(2l+1)2\pi^{2}}\,G_{n+3,l}} \lrb{-e^{-\alpha}\,\frac{2\pi^{2}}{T^{4}}\frac{G_{4,0} }{G_{5,0}G_{3,0}-G_{4,0}^{2}} } \\
    &=\frac{T^{n-1}}{(2l+1)}\frac{G_{n+4,l}G_{3,0} - G_{n+3,l}G_{4,0} }{G_{5,0}G_{3,0}-G_{4,0}^{2}}
\end{align}

\begin{align}
     \frac{\partial \rho_{n,l}^{eq}}{\partial \rm n} &=\frac{\partial \rho_{n,l}^{eq}}{\partial T} \frac{\partial T}{\partial \rm n} + \frac{\partial \rho_{n,l}^{eq}}{\partial \alpha} \frac{\partial \alpha}{\partial \rm n}\\
     &= \lrrb{e^{\alpha}\,\frac{T^{n+2}}{(2l+1)2\pi^{2}}G_{n+4,l}} ~ \lrb{-e^{-\alpha}\,\frac{2\pi^{2}}{T^{2}}\frac{G_{4,0} }{G_{5,0}G_{3,0}-G_{4,0}^{2}} } \\
     &~~~~~~~~~~~~+ \lrrb{e^{\alpha}\,\frac{T^{n+3}}{(2l+1)2\pi^{2}}\,G_{n+3,l}} ~ \lrb{e^{-\alpha}\,\frac{2\pi^{2}}{T^{3}}\frac{G_{5,0} }{G_{5,0}G_{3,0}-G_{4,0}^{2}} } \\
     &=\frac{T^{n}}{(2l+1)}\frac{ G_{5,0}G_{n+3,l} - G_{4,0}G_{n+4,l}}{G_{5,0}G_{3,0}-G_{4,0}^{2}}
\end{align}

so we find
\begin{align}\label{EqAp:cnlknl_explicit}
    c_{n,l}^{2} &= \frac{T^{n-1}}{(2l+1)}\frac{G_{n+4,l}G_{3,0} - G_{n+3,l}G_{4,0} }{G_{5,0}G_{3,0}-G_{4,0}^{2}} \,,\\
    \kappa_{n,l} &= {\rm n}\frac{T^{n}}{(2l+1)}\frac{ G_{5,0}G_{n+3,l} - G_{4,0}G_{n+4,l}}{G_{5,0}G_{3,0}-G_{4,0}^{2}}\,,.
\end{align}

As a check note that $\frac{\partial \rho_{1,0}^{eq}}{\partial \varepsilon} = c_{1,0}^{2} = 1$ and $\frac{\partial \rho_{0,0}^{eq}}{\partial \rm n} = \kappa_{0,0} = 1$. 

Finally, we get the speed of sound 
\begin{equation}
    c_{s}^{2}(z)  = c_{1,1}^{2}(z) = \frac{1}{3}\lrrb{\frac{G_{5,1}(z)G_{3,0}(z) - G_{4,1}(z)G_{4,0}(z) }{G_{5,0}(z) G_{3,0}(z) -G_{4,0}^{2}(z) }} \,.
\end{equation}

In the conformal limit, we have
\begin{align*}
    c_{s}^{2}(0) = \frac{1}{3}\lrrb{\frac{\Gamma(5) \Gamma(3) - \Gamma(4)\Gamma(4) }{\Gamma(5) \Gamma(3) - \Gamma(4)\Gamma(4) }} = \frac{1}{3}\,,.
\end{align*}

The bulk modulus,
\begin{align}
    \kappa(z) = \kappa_{1,1} = {\rm n}\frac{T}{(2l+1)}\frac{ G_{5,0}(z) G_{4,1}(z)  - G_{4,0}(z) G_{5,1}(z) }{G_{5,0}(z) G_{3,0}(z) -G_{4,0}^{2}(z) }\,,
\end{align}
which in the conformal limit
\begin{align*}
    \kappa(0) = {\rm n}\frac{T}{(2l+1)}\frac{ \Gamma(5)\Gamma(4) - \Gamma(4)\Gamma(5)}{\Gamma(5)\Gamma(3)-\Gamma(4)^{2}} = 0\,,
\end{align*}
vanishes.

\clearpage
\subsection{Evolution equations}

We start with computing the equilibrium gradient,
\begin{equation}
    \mathbf{\hat{D}} \rho^{\rm eq}_{n,l}(\tau) = \partial_{\tau}\rho^{\rm eq}_{n,l}(\tau) + \frac{(2l+1)}{\tau}\rho^{\rm eq}_{n,l}(\tau)- \frac{(2l-n)}{\tau}\rho^{\rm eq}_{n,l+1}(\tau)\,.
\end{equation}

We can express the time derivative term in terms of the energy derivative,
\begin{align}
    \partial_{\tau}\rho^{\rm eq}_{n,l}(\tau) &= \pdv{\rho^{\rm eq}_{n,l}}{\varepsilon} \pdv{\varepsilon}{\tau} + \pdv{\rho^{\rm eq}_{n,l}}{n}  \pdv{n}{\tau}\\
                                        &= -\pdv{\rho^{\rm eq}_{n,l}}{\varepsilon}~ \frac{(\varepsilon + P + \Pi-\pi)}{\tau}  - \pdv{\rho^{\rm eq}_{n,l}}{\text{n} }~ \frac{\text{n} }{\tau}\,.
\end{align}

Note that any $\mathbf{\hat{D}}^k \rho^{\rm eq}_{n,l}(\tau)$ can be reduced in terms of the hydrodynamic variables $\varepsilon$,$n$,$\Pi$ and $\pi$.

In the Bjorken case we can write,
\begin{equation}
    \partial_{\tau}\rho^{\rm eq}_{n,l}(\tau) = -c_{n,l}^2\frac{(\varepsilon + P + \Pi-\pi)}{\tau}- \frac{\kappa_{n,l} }{\tau}
\end{equation}

\begin{align}
    \mathbf{\hat{D}}\rho^{\rm eq}_{n,l}(\tau) &= -c_{n,l}^2\frac{(\varepsilon + P + \Pi-\pi)}{\tau}- \frac{\kappa_{n,l} }{\tau} + \frac{(2l+1)}{\tau}\rho^{\rm eq}_{n,l}(\tau)- \frac{(2l-n)}{\tau}\rho^{\rm eq}_{n,l+1}(\tau)\\
    &= -c_{n,l}^2 \frac{\pi_{1,1}}{\tau} - \lrrb{c_{n,l}^2(\varepsilon + P) + \kappa_{n,l} - \frac{3}{G_{n+3,1}(z)}\lrb{G_{n+3,l}(z) -\frac{(2l-n)}{(2l+3)}G_{n+3,l+1}(z)  }\rho_{n,1}^{\rm eq}  }\frac{1}{\tau}\,.
\end{align}
For the equilibrium pressure $\rho_{1,1}^{\rm eq}$then obtain, 
\begin{align}
     \mathbf{\hat{D}} P   &= -c_{s}^{2} \frac{(\varepsilon + P + \Pi-\pi  )}{\tau} - \frac{\kappa }{\tau} + \frac{3P}{\tau}- \frac{\rho^{\rm eq}_{1,2}(\tau)}{\tau} \\
     &= -c_{s}^{2} \frac{\pi_{1,1}}{\tau}-  \lrb{c_{s}^2(\varepsilon + P) + \kappa - 3\lrrb{1 -\frac{1}{5}\frac{G_{4,2}(z)}{G_{4,1}(z)}  }P  }\frac{1}{\tau}
\end{align}

The from \eqref{Eq:ApNEqEvol} we get the evolution equation for the anisotropic components,
\begin{equation}
    \mathbf{\hat{D}}\pi_{n,l}  = -\frac{\pi_{n,l} }{\tau_R} +c_{n,l}^2\frac{\pi_{1,1}}{\tau}+  \lrrb{ c_{n,l}^2(\varepsilon + P )+ \kappa_{n,l}  - \frac{3}{G_{n+3,1}(z)}\lrb{G_{n+3,l}(z) -\frac{(2l-n)}{(2l+3)}G_{n+3,l+1}(z)  }\rho_{n,1}^{\rm eq}  }~\frac{1}{\tau}\,.
\end{equation}

\subsection*{Evolution of $\Pi$}

The bulk pressure is related to the anisotropic moments $\Pi = -\frac{m^2}{3}\pi_{-1,0}$. The evolution equation for $\pi_{-1,0}$ is,

\begin{align}
     \mathbf{\hat{D}}\pi_{-1,0}  = -\frac{\pi_{-1,0} }{\tau_R}- \mathbf{\hat{D}} \rho^{\rm eq}_{-1,0}(\tau)
\end{align}

\begin{equation}
    \mathbf{\hat{D}}\pi_{-1,0}  = -\frac{\pi_{-1,0} }{\tau_R} +c_{-1,0}^2\frac{\pi_{1,1}}{\tau}+  \lrrb{ c_{-1,0}^2(\varepsilon + P )+ \kappa_{-1,0}  - \frac{3}{G_{2,1}(z)}\lrb{G_{2,0}(z) -\frac{(2\cdot 0 +1)}{(2\cdot 0+3)}G_{2,1}(z)  }\rho_{-1,1}^{\rm eq}  }~\frac{1}{\tau}\,.
\end{equation}

Multiplying both sides by $-m^2/3$, we get the equation for $\Pi$,
\begin{align*}
    \mathbf{\hat{D}}\Pi  &= -\frac{\Pi }{\tau_R} - \frac{m^2}{3}c_{-1,0}^2\frac{\pi_{1,1}}{\tau}-  \lrrb{ \frac{m^2}{3}c_{-1,0}^2(\varepsilon + P )+ \frac{m^2}{3}\kappa_{-1,0}  - \frac{3}{G_{2,1}(z)}\lrb{G_{2,0}(z) -\frac{(2\cdot 0 +1)}{(2\cdot 0+3)}G_{2,1}(z)  }(\frac{m^2}{3}\rho_{-1,1}^{\rm eq})  }~\frac{1}{\tau} \nonumber\\
    &= -\frac{\Pi }{\tau_R} - \frac{m^2}{3}c_{-1,0}^2\frac{\pi_{1,1}}{\tau}-  \lrrb{ \frac{m^2}{3}c_{-1,0}^2 (\varepsilon + P )+ \frac{m^2}{3}\kappa_{-1,0}  - \lrb{ \frac{G_{2,0}(z)}{G_{2,1}(z)} -\frac{1}{3} }\frac{G_{2,1}}{3G_{2,0}}(\varepsilon-3P)  }~\frac{1}{\tau}\\
    &= -\frac{\Pi }{\tau_R} - \frac{m^2}{3}c_{-1,0}^2\frac{\pi_{1,1}}{\tau}-  \lrrb{ \frac{m^2}{3}c_{-1,0}^2 (\varepsilon + P )+ \frac{m^2}{3}\kappa_{-1,0}  - \frac{1}{3}\lrb{ 1 -\frac{1}{3}\frac{G_{2,0}(z)}{G_{2,1}(z)} }(\varepsilon-3P)  }~\frac{1}{\tau}
\end{align*}

We now relate $\frac{m^2}{3}c_{-1,0}^2$ to $c_{s}^2 = c_{1,1}^2$ using the relations from Eq\eqref{EqAp:cnlknl_explicit} and Eq.\eqref{EqAp:z2Gnl}\,. We have the explicit expression,
\begin{align}
    \frac{m^2}{3}c_{-1,0}^2 &= \frac{m^2}{3}~\frac{T^{-1-1}}{(2\cdot 0+1)}\frac{G_{-1+4,l}G_{3,0} - G_{-1+3,0}G_{4,0} }{G_{5,0}G_{3,0}-G_{4,0}^{2}} \nonumber\\
    &=\frac{1}{3}\frac{m^2}{T^2}~\frac{G_{3,0}G_{3,0} - G_{2,0}G_{4,0} }{G_{5,0}G_{3,0}-G_{4,0}^{2}} \nonumber\\
    &= \frac{1}{3}\frac{ (z^2 G_{3,0})G_{3,0} - (z^2G_{2,0})G_{4,0} }{G_{5,0}G_{3,0}-G_{4,0}^{2}}\\
    &= \frac{1}{3}\frac{ ( G_{5,0} - G_{5,1})G_{3,0} - (G_{4,0}-G_{4,1})G_{4,0} }{G_{5,0}G_{3,0}-G_{4,0}^{2}} \nonumber\\
    &=\frac{1}{3} \frac{ (G_{5,0}G_{3,0}-G_{4,0}^{2}) - (G_{5,1}G_{3,0}-G_{4,1}G_{4,0}) }{G_{5,0}G_{3,0}-G_{4,0}^{2}} \nonumber\\
    &=\frac{1}{3}\lrrb{ 1 -\frac{G_{5,1}G_{3,0}-G_{4,1}G_{4,0}}{G_{5,0}G_{3,0}-G_{4,0}^{2}} }\\
    &= \frac{1}{3} - c_{s}^{2}\,.
\end{align}

Similarly for $\frac{m^2}{3}\kappa_{-1,0}$ we have
\begin{align}
    \frac{m^2}{3}\kappa_{-1,0}^2 &= \frac{m^2}{3}~ {\rm n}\frac{T^{-1}}{(2\cdot 0+1)}\frac{ G_{5,0}G_{-1+3,0} - G_{4,0}G_{-1+4,0}}{G_{5,0}G_{3,0}-G_{4,0}^{2}} \nonumber\\
    &={\rm n}\frac{T}{3}\frac{ G_{5,0}(G_{4,0}-G_{4,1} )- G_{4,0}(G_{5,0} - G_{5,1})   }{ G_{5,0}G_{3,0}-G_{4,0}^{2}} \nonumber\\
    &=-{\rm n}\frac{T}{3} \frac{G_{5,0}G_{4,1} - G_{5,1}G_{4,0}}{ G_{5,0}G_{3,0}-G_{4,0}^{2}} = -\kappa
\end{align}

From this we get
\begin{align}
    \mathbf{\hat{D}}\Pi = -\frac{\Pi }{\tau_R} - \lrrb{\frac{1}{3} - c_{s}^{2}}\frac{\pi_{1,1}}{\tau}- \lrrb{\lrb{\frac{1}{3} - c_{s}^{2}} (\varepsilon + P) - \kappa - \frac{1}{3}\lrb{1 -\frac{1}{3} \frac{G_{2,1}(z)}{G_{2,0}(z)}  }(\varepsilon-3P)  }\frac{1}{\tau}
\end{align}

Using
\begin{equation}
    \mathbf{\hat{D}}\Pi = \partial_{\tau} \Pi + \frac{1}{\tau}\Pi +\frac{m^2}{3} \frac{1}{\tau}\pi_{-1,1}
\end{equation}
we get the final expression,
\begin{align}
    \partial_{\tau} \Pi+ \frac{\Pi }{\tau_R}= -\lrrb{\frac{4}{3} - c_{s}^{2}}\frac{\Pi}{\tau} + \lrrb{\frac{1}{3} - c_{s}^{2}}\frac{\pi}{\tau} - \lrrb{\lrb{\frac{1}{3} - c_{s}^{2}} (\varepsilon + P) - \kappa - \frac{1}{3}\lrb{1 -\frac{1}{3} \frac{G_{2,1}(z)}{G_{2,0}(z)}  }(\varepsilon-3P)  }\frac{1}{\tau} - \frac{m^2}{3} \frac{1}{\tau}\pi_{-1,1}
\end{align}

\subsection*{Evolution of $\pi$}
The evolution equation for $\pi_{1,1}$ reads,
\begin{align}
     \mathbf{\hat{D}}\pi_{1,1} &= -\frac{\pi_{1,1} }{\tau_R} + c_{s}^{2}\frac{\pi_{1,1}}{\tau} + \lrrb{c_{s}^{2} (\varepsilon + P ) + \kappa  - 3\lrb{ 1- \frac{1}{5}~\frac{G_{4,2}}{G_{4,1}}}P  } \frac{1}{\tau}\,.
\end{align}    
Now,
\begin{align}
    \mathbf{\hat{D}}\pi_{1,1} &= \partial_{\tau}\pi_{1,1} + (2\cdot 1 + 1)\frac{\pi_{1,1} }{\tau} -  (2\cdot 1 - 1)\frac{\pi_{1,2} }{\tau}\\
    &= \partial_{\tau}\pi_{1,1} + 3\frac{\pi_{1,1} }{\tau} -  \frac{\pi_{1,2} }{\tau}
\end{align}

Using this, we get the evolution equation for $\pi_{1,1}$,
\begin{align}\label{Eq:Evolpi11}
    \partial_{\tau}\pi_{1,1} + \frac{\pi_{1,1} }{\tau_R} =  -( 3- c_{s}^{2})\frac{\pi_{1,1}}{\tau}  + \lrrb{c_{s}^{2} (\varepsilon + P ) + \kappa  - 3\lrb{ 1 - \frac{1}{5}\frac{G_{4,2}(z)}{G_{4,1}(z)} }P  } \frac{1}{\tau}   + \frac{\pi_{1,2}}{\tau} 
\end{align}

Now from the equation \eqref{Eq:Evolpi11},
\begin{align}
    \partial_{\tau}(\pi-\Pi) + \frac{(\pi-\Pi) }{\tau_R} =  -( 3- c_{s}^{2})\frac{(\pi-\Pi)}{\tau}  - \lrrb{c_{s}^{2} (\varepsilon + P ) + \kappa  - 3\lrb{ 1- \frac{1}{5}\frac{G_{4,2}(z)}{G_{4,1}(z)} }  } \frac{1}{\tau}   - \frac{\pi_{1,2}}{\tau} 
\end{align}

\begin{align}
    \partial_{\tau}\pi + \frac{\pi }{\tau_R} &= \lrb{\partial_{\tau}\Pi + \frac{\Pi }{\tau_R}}\\
                        &~~~~~~~~~~~~-( 3- c_{s}^{2})\frac{(\pi-\Pi)}{\tau}  - \lrrb{c_{s}^{2} (\varepsilon + P ) + \frac{\kappa }{\tau} - 3\lrb{ 1- \frac{1}{5}\frac{G_{4,2}(z)}{G_{4,1}(z)} } P  } \frac{1}{\tau}   - \frac{\pi_{1,2}}{\tau} \nonumber\\
                        &=  -\lrrb{\frac{4}{3} - c_{s}^{2}}\frac{\Pi}{\tau} + \lrrb{\frac{1}{3} - c_{s}^{2}}\frac{\pi}{\tau} - \lrrb{\lrb{\frac{1}{3} - c_{s}^{2}} (\varepsilon + P) - \kappa - \frac{1}{3}\lrb{1 -\frac{1}{3} \frac{G_{2,1}(z)}{G_{2,0}(z)}  }(\varepsilon-3P)  }\frac{1}{\tau} - \frac{m^2}{3} \frac{1}{\tau}\pi_{-1,1} \nonumber\\
                        &~~~~~~~~~~~~~~~-( 3- c_{s}^{2})\frac{(\pi-\Pi)}{\tau}  - \lrrb{c_{s}^{2} (\varepsilon + P ) + \kappa  -  3\lrb{ 1- \frac{1}{5}\frac{G_{4,2}(z)}{G_{4,1}(z)} } P  } \frac{1}{\tau}   - \frac{\pi_{1,2}}{\tau} \nonumber\\
                        &=- \frac{8}{3\tau}\pi + \frac{5}{3\tau}\Pi - \lrrb{\frac{1}{3}(\varepsilon + P) -\frac{1}{3}\lrb{1 -\frac{1}{3} \frac{G_{2,1}(z)}{G_{2,0}(z)}  }(\varepsilon -3P)- 3\lrb{ 1 -\frac{1}{5}\frac{G_{4,2}(z)}{G_{4,1}(z)} }P}\frac{1}{\tau}   -\frac{m^2}{3}\frac{\pi_{-1,1} }{\tau}-\frac{\pi_{1,2}}{\tau}
\end{align}

Interestingly for the Conformal case, we have the relation,
\begin{align}
    \partial_{\tau}\pi + \frac{\pi }{\tau_R} &= -( 3- c_{s}^{2})\frac{\pi}{\tau} - \lrrb{c_{s}^{2} (\varepsilon + P ) + \frac{\kappa }{\tau} - 3\lrb{ 1- \frac{1}{5}\frac{G_{4,2}(0)}{G_{4,1}(0)} }P  }\frac{1}{\tau} -\frac{\pi_{1,2}}{\tau}\\
    &=-\frac{8}{3}\frac{\pi}{\tau}- \lrrb{ \frac{4}{3}P +  0 - \frac{3\cdot4}{5}P}\frac{1}{\tau}  -\frac{\pi_{1,2}}{\tau}\\
    &=-\frac{8}{3}\frac{\pi}{\tau} +\frac{4}{3} \lrb{\frac{4}{5}P}\frac{1}{\tau}-\frac{\pi_{1,2}}{\tau}
\end{align}

\subsection{Hydrodynamic equations}
Collecting all the results we have for a massive system, the equations,
\begin{align}
  \partial_{\tau}\pi + \frac{\pi }{\tau_R} &= - \frac{8}{3\tau}\pi + \frac{5}{3\tau}\Pi - \lrrb{\frac{1}{3}(\varepsilon + P) - 3\lrb{1-\frac{1}{5}\frac{G_{4,2}(z)}{G_{4,1}(z)}}P-  \frac{1}{3}\lrb{1 -\frac{1}{3} \frac{G_{2,1}(z)}{G_{2,0}(z)}  }(\varepsilon-3P)   }\frac{1}{\tau}   -\frac{m^2}{3}\frac{\pi_{-1,1} }{\tau}-\frac{\pi_{1,2}}{\tau}\\
  \partial_{\tau} \Pi+ \frac{\Pi }{\tau_R}&= -\lrrb{\frac{4}{3} - c_{s}^{2}}\frac{\Pi}{\tau} + \lrrb{\frac{1}{3} - c_{s}^{2}}\pi- \lrrb{\lrb{\frac{1}{3} - c_{s}^{2}} (\varepsilon + P) - \kappa - \frac{1}{3}\lrb{1 -\frac{1}{3} \frac{G_{2,1}(z)}{G_{2,0}(z)}  }(\varepsilon-3P)  }\frac{1}{\tau} - \frac{m^2}{3} \frac{1}{\tau}\pi_{-1,1}\,.
\end{align}

%%%------------------------------------------------------
\section{$3~+~1$D Calculations}

The $3~+~1$D moment equations have the form,
\begin{equation}
    \lrrb{\partial_{\tau} + \hat{\textbf{F}} +  \hat{\boldsymbol{\nu}}_{R} }\vec{\rho} =  \hat{\boldsymbol{\nu}}_{R} \vec{\rho}^{~\text{eq}}\,.
\end{equation}

%%{\color{Red}Explain this structure}

Our goal is to construct the propagator $U(\tau,\tau_0)$ for this linear equation. To do so, we employ perturbative techniques familiar from linear differential equations such as the Schrödinger equation, using the interaction picture. In this approach, the operator is decomposed into an exactly solvable part and an interaction term, and the evolution is then expressed in a basis governed by the solvable dynamics.

Here, two operators govern the evolution: the free-streaming Liouville operator $\hat{\textbf{L}}$ and the collision operator $\hat{\boldsymbol{\nu}}_{R}$. Either of them may be designated as the interaction, depending on which choice yields the most convenient form for the final expression.

\subsection{Collision picture}\label{Ap:ThpOColP}

 We will choose to decompose the propagator into free streaming and collision part and treat the collision as the interaction.  Consider the free streaming equation, 
 \begin{equation}
      \lrrb{\partial_{\tau} + \hat{\mathbf{F}} } = 0
 \end{equation}

 with the propagator $\hat{\mathcal{F}}$ satisfying,
\begin{equation}
   \partial_{\tau}\hat{\mathcal{F}}_{\tau,\tau_0}  = -\hat{\mathbf{F}}~\hat{\mathcal{F}}_{\tau,\tau_0}
\end{equation}

We can now define the collision picture operator and variables,
\begin{equation}
    \vec{\rho}_{c}(\tau) = \hat{\mathcal{F}}^{-1}_{\tau,\tau_0}\vec{\rho}(\tau)
\end{equation}
and 
\begin{equation}
   \hat{\boldsymbol{\nu}}_{c}  = \hat{\mathcal{F}}^{-1}_{\tau,\tau_0}~\hat{\boldsymbol{\nu}}_{R}~\hat{\mathcal{F}}_{\tau,\tau_0}
\end{equation}
Our collision picture with the interaction term equation now reads
\begin{equation}
    \lrrb{\partial_{\tau}  + \hat{\boldsymbol{\nu}}_{c} }~\vec{\rho}_{c} =0
\end{equation}
The propagator for the interaction picture is given by
\begin{equation}
    \mathcal{W}_{\tau,\tau_0} = \mathcal{T}\lrb{e^{-\int_{\tau_0}^{\tau}d\tau'\hat{\boldsymbol{\nu}}_{c}}}
\end{equation}
where $\mathcal{T}$ represents the time ordering. The solution to the interaction picture equation is
\begin{equation}
    \vec{\rho}_{c}{(\tau)} =  \mathcal{W}_{\tau,\tau_0} \rho_{c}{(\tau_0)} + \int_{\tau_0}^{\tau} d\tau' \mathcal{W}_{\tau,\tau'}~\hat{\boldsymbol{\nu}}_{c} (\tau')\vec{\rho}_{c}^{~\text{eq}}(\tau')
\end{equation}

We can therefore write,
\begin{align}
    \int_{\tau_0}^{\tau} d\tau'\mathcal{W}_{\tau,\tau'}~\hat{\boldsymbol{\nu}}_{c} (\tau')\vec{\rho}_{c}^{~\text{eq}}(\tau') &= \int_{\tau_0}^{\tau} d\tau'\partial_{\tau'} \mathcal{W}_{\tau,\tau'}\rho_{c}^{~\text{eq}}(\tau') \nonumber\\
    &= \mathcal{W}_{\tau,\tau'}\rho_{c}^{~\text{eq}}(\tau')\Bigg|_{\tau_0}^{\tau} -  \int_{\tau_0}^{\tau} d\tau' \mathcal{W}_{\tau,\tau'}\partial_{\tau'}\rho_{c}^{~\text{eq}}(\tau') \nonumber\\
    &= \mathcal{W}_{\tau,\tau'}\rho_{c}^{~\text{eq}}(\tau')\Bigg|_{\tau_0}^{\tau} -  \int_{\tau_0}^{\tau} d\tau' \mathcal{W}_{\tau,\tau'}\hat{\boldsymbol{\nu}}_{c}(\tau') \hat{\boldsymbol{\nu}}_{c}^{-1}(\tau')\partial_{\tau'}\rho_{c}^{~\text{eq}}(\tau') \nonumber\\
    &=\sum_{n=0}^{\infty} \mathcal{W}_{\tau,\tau'}~  \lrrb{\hat{\boldsymbol{\nu}}_{c}^{-1}(\tau')\partial_{\tau'}}^{k}~\rho_{c}^{~\text{eq}}(\tau')\Bigg|_{\tau_0}^{\tau} \nonumber\\
    &=\hat{\mathcal{F}}^{-1}_{\tau,\tau_0}\sum_{n=0}^{\infty}  \lrrb{\hat{\boldsymbol{\nu}}^{-1}_{R}(\tau)\mathbf{\hat{D}}}^{k}~\rho^{~\text{eq}}(\tau) - \mathcal{W}_{\tau,\tau_0} ~\sum_{n=0}^{\infty}  \lrrb{\hat{\boldsymbol{\nu}}^{-1}_{R}(\tau_0)\mathbf{\hat{D}}}^{k}~\rho^{~\text{eq}}(\tau_0)
\end{align}

where in the last step we used the collision picture definitions. Now we get the solution in the Schrodinger picture
\begin{equation}
   \vec{\rho}(\tau) = \hat{\mathcal{F}}_{\tau,\tau_0}\mathcal{W}_{\tau,\tau_0} \rho(\tau_0)  + \sum_{n=0}^{\infty}  \lrrb{\hat{\boldsymbol{\nu}}^{-1}_{R}(\tau)\mathbf{\hat{D}}}^{k}~\rho^{~\text{eq}}(\tau) - \hat{\mathcal{F}}_{\tau,\tau_0}\mathcal{W}_{\tau,\tau_0} ~\sum_{n=0}^{\infty}  \lrrb{\hat{\boldsymbol{\nu}}^{-1}_{R}(\tau_0)\mathbf{\hat{D}}}^{k}~\rho^{~\text{eq}}(\tau_0)
\end{equation}

%%------------------------------------------------------------------------------
\subsection{Free Streaming picture}\label{Ap:ThpOFSP}

To derive the gradient expansion in the $3~+~1$D case, we adopt an alternative approach in which free streaming is treated as a perturbation. Consider the damping equation,
\begin{equation}
    \lrrb{\partial_{\tau} + \hat{\boldsymbol{\nu}}_{R}  }\vec{\rho}(\tau) = 0\,.
\end{equation}
with the associated propagator obeying,
\begin{equation}
    \partial_{\tau}\mathcal{W}_{\tau,\tau_0} = -\hat{\boldsymbol{\nu}}_{R} ~ \mathcal{W}_{\tau,\tau_0}\,. 
\end{equation}
The propagator has the simple form,
\begin{equation}
     \mathcal{W}_{\tau,\tau_0}  = e^{-\int_{\tau_0}^{\tau}d\tau''\frac{1}{\tau_R}}\hat{I}\,.
\end{equation}

We now define the streaming–picture field and operator as,
\begin{equation}
    \rho_{S}(\tau)= \mathcal{W}_{\tau,\tau_0}\rho(\tau)
\end{equation}
and operator,
\begin{equation}
    \hat{\textbf{L}}_{S} = \mathcal{W}_{\tau,\tau_0}\hat{\textbf{L}}\mathcal{W}^{-1}_{\tau,\tau_0}
\end{equation}

We then have the solution,
\begin{equation}
    \lrrb{\partial_{\tau} + \hat{\mathbf{F}}_{S}}\vec{\rho}_{S} = \partial_{\tau}\vec{\rho}_{S}
\end{equation}

The corresponding free–streaming picture propagator is,
\begin{equation}
    \hat{\mathcal{F}}_{\tau,\tau_0} = \mathcal{T}e^{-\int_{\tau_0}^{\tau} d\tau\hat{\textbf{F}}_{S}}
\end{equation}
which satisfies,
\begin{equation}
    \partial_{\tau}\hat{\mathcal{F}}_{\tau,\tau_0} =  -\hat{\textbf{F}}_{S}~\hat{\mathcal{F}}_{\tau,\tau_0} 
\end{equation}

The full solution can then be expressed as,
\begin{equation}
    \rho(\tau) = e^{-\xi_0}\hat{\mathcal{F}}_{\tau,\tau_0}\rho(\tau_0) + \int_{\tau_0}^{\tau} d\tau'e^{-\xi'}\hat{\mathcal{F}}_{\tau,\tau'} \frac{\vec{\rho}^{~\text{eq}}(\tau')}{\tau_R}
\end{equation}
Since $\xi$ does not commute with $\hat{\mathcal{F}}{\tau,\tau’}$, it cannot be treated as a simple scalar integration variable. Instead, we expand $\hat{\mathcal{F}}_{\tau,\tau'} \frac{\vec{\rho}^{~\text{eq}}(\tau')}{\tau_R}$ in a Taylor series,
\begin{align}
    \hat{\mathcal{F}}_{\tau,\tau'} \frac{\vec{\rho}^{~\text{eq}}(\tau')}{\tau_R} &= \sum_{n=0}^{\infty} \frac{(t-t')^n}{n!}\partial_{\tau'}^n\lrb{\hat{\mathcal{F}}_{\tau,\tau'} \frac{\vec{\rho}^{~\text{eq}}(\tau')}{\tau_R}} \Bigg|_{\tau'=\tau} \nonumber\\
    &=\sum_{n=0}^{\infty} \frac{(t-t')^n}{n!} \lrrb{\partial_{\tau} + \mathbf{\hat{F}} }^{n} \frac{\vec{\rho}^{~\text{eq}}(\tau')}{\tau_R} \,.
\end{align}

Define the function 
\begin{equation}
    g_n(\tau,\tau_0) = \int_{\tau_0}^{\tau} d\tau'e^{-\xi'} (t-t')^n 
\end{equation}

The hydrodynamic generator can be then expressed in terms of the gradients as,
\begin{align}
   \vec{\rho}^{\rm ~G}(\tau) &= \int_{\tau_0}^{\tau} d\tau'e^{-\xi'}\hat{\mathcal{F}}_{\tau,\tau'} \frac{\rho^{~\text{eq}}(\tau')}{\tau_R} \\
   &= \sum_{n=0}^{\infty}\frac{g_n(\tau,\tau_0)}{n!} ~ \lrrb{\partial_{\tau} + \mathbf{\hat{F}} }^{n} \frac{\vec{\rho}^{~\text{eq}}(\tau')}{\tau_R}\,.
\end{align}

Although this expansion does not take exactly the same form as in the commuting case, it can still be reorganized in terms of gradient operators appearing in the gradient expansion, even though a compact closed-form expression remains out of reach. Note that

\begin{align}
    \lrrb{\partial_{\tau} + \mathbf{\hat{F}} }^{n} \frac{\vec{\rho}^{~\text{eq}}(\tau')}{\tau_R} & =  (-1)^n\lrrb{\frac{-\tau_R}{\tau_R}\lrb{\partial_{\tau} + \mathbf{\hat{F}} }}^{n} \frac{ \vec{\rho}^{~\text{eq}}(\tau')}{\tau_R}\\
    &= \sum_{k=0}^{n}h_{n,k}(-\tau_R,\mathbf{\hat{D}} \tau_R)  \lrrb{\tau_R \mathbf{\hat{D}}  }^{n}\vec{\rho}^{~\text{eq}}
\end{align}
where $h_{n,k}(\tau_R,\mathbf{\hat{D}})$ is a function of $\tau_R$ and its derivatives $\mathbf{\hat{D}}\tau_R$. We then get the copact expression,
\begin{align}
    \vec{\rho}^{~\rm G}(\tau) &=  \sum_{n=0}^{\infty}\frac{g_n(\tau,\tau_0)}{n!}\sum_{k=0}^{n}h_{n,k}(\tau_R,\mathbf{\hat{D}} \tau_R)  \lrrb{-\tau_R \mathbf{\hat{D}}  }^{k}\vec{\rho}^{~\text{eq}}\\
    &=  \sum_{n=0}^{\infty}~\lrb{\sum_{k=n}^{\infty} \frac{g_k(\tau,\tau_0)}{n!}h_{k,n}(\tau_R,\mathbf{\hat{D}} \tau_R) }~\lrrb{-\tau_R \mathbf{\hat{D}}  }^{n}\vec{\rho}^{~\text{eq}}
\end{align}

\subsection{Equilibrium gradients and closure}\label{Ap:EqGradClos}

This section describe the standard formalism used to extract transport coefficients and higher order gradient expansion.

Consider the gradients of the equilibrium moments,  $\lrrb{-\hat{\boldsymbol{\nu}}^{-1}_{R}   \mathbf{\hat{D}}}^{n}\vec{\rho}^{~\text{eq}}(\tau)$, where
\begin{align}
    \mathbf{\hat{D}} \equiv \lrrb{\partial_{\tau} + \hat{\mathbf{F}}_{\lambda}\nabla^{\lambda} +\hat{\mathbf{A}} }\,.
\end{align}
As $\vec{\rho}^{~\text{eq}}(\tau)$ is a function of $T,~\mu,~u^{\lambda}$, we can trade the space and time derivatives of $\vec{\rho}$ with space time derivatives of these hydrodynamic quantities.
\begin{align}
    \partial_{\tau}\vec{\rho}^{~\text{eq}} &= \partial_{T}\vec{\rho}^{~\text{eq}}~\partial_{\tau}T + \partial_{\mu}\vec{\rho}^{~\text{eq}}~\partial_{\tau}\mu +\partial_{u^{\lambda}}\vec{\rho}^{~\text{eq}}~\partial_{\tau}u^{\lambda}\\
    \nabla_{\sigma}\vec{\rho}^{~\text{eq}} &= \partial_{T}\vec{\rho}^{~\text{eq}}~\nabla_{\sigma}T + \partial_{\mu}\vec{\rho}^{~\text{eq}}~\nabla_{\sigma}\mu +\partial_{u^{\mu}}\vec{\rho}^{~\text{eq}}~\nabla_{\sigma}u^{\mu}
\end{align}

We can now use the conservation equations,
\begin{align}
\partial_{\tau} n + n\,\partial_{\mu} u^{\mu}+ \partial_{\mu} n^{\mu}&= 0, \\[6pt]
\partial_{\tau} \epsilon+ (\epsilon + P + \Pi)\,\partial_{\mu} u^{\mu}-\pi^{\mu\nu}\,\nabla_{(\mu}u_{\nu)}&= 0, \\[6pt]
(\epsilon + P + \Pi)\,\partial_{\tau} u^{\alpha}- \nabla^{\alpha}(P + \Pi)+ \Delta^{\alpha}{}_{\nu}\,\partial_{\mu}\pi^{\mu\nu}&= 0 .
\end{align}
and the equation of state, $\epsilon = \epsilon (T,\mu),~ n =n(T,\mu),~P =P(T,\mu)$ to eliminate the derivatives of $T$ and $\mu$ in favor of gradients of $u^{\lambda}$ and the anisotropic moments $\Pi$, $\pi^{\mu\nu}$. Now any gradient of $\Pi$ or $\pi^{\mu\nu}$ appear in the next order in expansion. These can be replaced with the expressions found in the lower order expansion. In the scheme presented here, there is a crucial difference that this includes the transient corrections.

This procedure reduces all order gradients in term of the gradients of $u^{\mu}$ and the other hydrodynamic quantities $\varepsilon, n,P, \Pi, \pi^{\mu\nu}$. In general we can write
\begin{align}
    \sum_{k=0}^{n}\lrrb{-\hat{\boldsymbol{\nu}}^{-1}_{R}   \mathbf{\hat{D}}}^{k}\vec{\rho}^{~\text{eq}}(\tau) =  \vec{\rm a}^{~(n)}~ \epsilon + \vec{b}^{~(n)}~{\rm n} + \vec{\rm c}^{~(n)}~\Pi + \vec{d}_{\lambda\sigma}^{~(n)}~\pi^{\lambda\sigma}
\end{align}

Where the coefficients are functions of $\varepsilon, n,P, \partial_{\lambda}u^{\sigma}$. Consider the projection of the above equation to the hydrodynamic subspace,
\begin{align}
     \lrb{\vec{\rm a}^{~(n)}_{\mathcal{H}}~ \varepsilon + \vec{b}^{~(n)}_{\mathcal{H}}~{\rm n} + \vec{\rm c}^{~(n)}_{\mathcal{H}}~\Pi + \vec{d}_{\mathcal{H}~\lambda\sigma}^{~(n)}~\pi^{\lambda\sigma}}_{i} &= (\vec{\rm a}^{~(n)}_{\mathcal{H}})_{i} \varepsilon + (\vec{b}^{~(n)}_{\mathcal{H}})_{i}~{\rm n} + (\vec{\rm c}^{~(n)}_{\mathcal{H}})_{i}~\Pi + (\vec{d}_{\mathcal{H}~\lambda\sigma}^{~(n)})_{i}~\pi^{\lambda\sigma} \nonumber\\
     &= \sum_{h=0}^{N}( \mathbf{\hat{\Delta}}_{i,j})_{h}(\vec{\pi}_{\mathcal{H}})_{h}
\end{align}
where $ \mathbf{\hat{\Delta}}_{i,j}$ is an $N\times N$ matrix and we defined,
\begin{align}
    \vec{\pi}_{\mathcal{H}} = \begin{bmatrix}
        \epsilon\\
        \rm n\\
        \Pi\\
        \pi^{0,0}\\
        \pi^{0,1}\\
        \vdots\\\pi^{4,4}
    \end{bmatrix}
\end{align}
containing the fourteen independent hydrodynamic variables and 
\begin{align}
   \mathbf{\hat{\Delta}}= \begin{bmatrix}
        (\vec{a}){0} & (\vec{b}){0} &(\vec{c}){0}& (\vec{d}_{1,0})_{0} &\dots&(\vec{d}_{4,4})_{0}\\
        (\vec{a}){1} & (\vec{b}){1} &(\vec{c}){1}& (\vec{d}_{1,0})_{1} &\dots&(\vec{d}_{4,4})_{1}\\
        (\vec{a}){2} & (\vec{b}){2} &(\vec{c}){2}& (\vec{d}_{1,0})_{2} &\dots&(\vec{d}_{4,4})_{2}\\
        (\vec{a}){3} & (\vec{b}){3} &(\vec{c}){3}& (\vec{d}_{1,0})_{3} &\dots&(\vec{d}_{4,4})_{3}\\
        \vdots      & \vdots        & \vdots & \vdots    &\dots& \vdots \\
        (\vec{a}){14} & (\vec{b})_{14} & (\vec{c})_{14}& (\vec{d}_{1,0})_{14} &\dots & (\vec{d}_{4,4})_{14}
    \end{bmatrix}
\end{align}
is a $14\times 14$ matrix corresponding to the the fourteen independent hydrodynamic variables.

\newpage
 % your .bib file here
\bibliographystyle{apsrev4-2}
\bibliography{main}

\end{document}